\documentclass[smallextended]{svjour3}
\smartqed
\usepackage[applemac]{inputenc}
\usepackage{amsmath,amssymb}
\usepackage{graphicx}
\begin{document}
\title{Multidomain Galerkin-Collocation method:}
\subtitle{characteristic spherical collapse of scalar fields}
\author{M. A. Alcoforado$^{1}$ \and W. O. Barreto$^{1,\,2}$ \and \newline H. P. de Oliveira$^{1}$}
\institute{$^{1}\,\,\,${Departamento de F\'{\i}sica Te\'orica - 
Instituto de F\'{\i}sica A. D. Tavares, Universidade do Estado  do Rio de Janeiro,  
R. S\~ao Francisco Xavier, 524. Rio de Janeiro, RJ, 20550-013, Brazil}\\
$^{2}\,\,\,${Centro de F\'{\i}sica Fundamental, Universidad de Los Andes, M\'erida 5101,  Venezuela}}
\date{\today}
\maketitle
\begin{abstract}
We initiate a systematic implementation of the spectral domain decomposition technique with the Galerkin-Collocation (GC) method in situations of interest such as the spherical collapse of a scalar field in the characteristic formulation. We discuss the transmission conditions at the interface of contiguous subdomains that are crucial for the domain decomposition technique for hyperbolic problems. We implemented codes with an arbitrary number of subdomains, and after validating them, we applied to the problem of critical collapse. With a modest resolution, we obtain the Choptuik's scaling law and its oscillatory component due to the discrete self-similarity of the critical solution.  
\end{abstract}

\tableofcontents

%%%%%%%%%%%%%%%%%%%%%%
\section{Introduction}
%%%%%%%%%%%%%%%%%%%%%%

Spectral methods have become popular in numerical relativity with applications in a large variety of problems \cite{grand_novak}. It is well known that spectral methods are global methods characterized by expansion of high order polynomial approximations that provide highly accurate solutions exhibiting exponential convergence for smooth functions with moderate computational resources. However, in general, the accuracy is spoiled in the case of complex geometries, or for the situation in which the solutions have localized regions of rapid variation \cite{canuto_88,gottlieb_01}.

The domain decomposition or multidomain method is a powerful tool to deal with problems described by partial differential equations in complex geometries or presenting strong gradients in localized regions of the spatial domain. The multidomain technique consists of partitioning the physical domain into several subdomains together with the transmission conditions to connect the solutions in all subdomains. In fact, how solutions of contiguous subdomains are matched turn to be a crucial aspect of any domain decomposition scheme.

The spectral-domain decomposition codes were first established in problems of fluid mechanics in the late 1970s. We indicate Canuto et al. \cite{canuto_88} (see also Ref. \cite{canuto_new}) for a helpful and concise presentation of domain decomposition methods. Orszag \cite{orszag_80} introduced the spectral domain decomposition method for elliptic problems; Kopriva \cite{kopriva_86,kopriva_89} considered the spectral multidomain technique for hyperbolic problems. In this case, we remark that there is no unique way of matching the solutions in contiguous subdomains \cite{kopriva_86,faccioli_96}.

The first application of the spectral domain decomposition method in Numerical Relativity was to determine the stationary configurations \cite{bona} and the initial data problem \cite{pfeifer,ansorg}. For the time-dependent systems, the spectral domain decomposition was implemented within the SpEC \cite{spec} and LORENE \cite{lorene} codes to deal with the gravitational collapse, the dynamics of stars and the evolution of single \cite{kidder_00} and binary black holes \cite{szilagyi_09}. A more detailed approach for the multidomain spectral codes is found in Refs. \cite{Hemberger_13} and in the SXS collaboration \cite{sxs_col}.

In this work, we begin a systematic implementation of the multidomain technique with the Galerkin-Collocation method. As the first step, we consider the dynamics of self-gravitating spherically symmetric scalar fields described in the characteristic scheme \cite{winicour_12}. We have already implemented Galerkin-Collocation codes successfully in two domains in the following cases: (i) determination of the initial data for single \cite{oliveira_14} and binary black holes \cite{barreto_18,barreto_18_2}, (ii) the dynamics of cylindrical gravitational waves \cite{barreto_19}, and (iii) critical collapse in the new characteristic scheme \cite{crespo_19}.

We divided the paper as follows. In the second Section, we present the field equations and the essential elements of the characteristic scheme. We describe the domain-decomposition strategy in Section 3. We remark the innovative introduction of the computational subdomains, the basis functions' construction, and the distribution of collocation points in each subdomain. We have also briefly discussed the transmission conditions for the present hyperbolic problem.  In Section 4, we present the numerical tests to validate the multidomain code. We have applied the code to reproduce the Choptuik scaling law and its oscillatory component in connection with the critical collapse. We summarize the results and trace out some directions of the present multidomain strategy.

%%%%%%%%%%%%%%%%%%%%%%%%%%%%%%
\section{The field equations}%
%%%%%%%%%%%%%%%%%%%%%%%%%%%%%%

We present briefly the main equations and aspects of the spherical collapse of a scalar field described by the characteristic scheme \cite{winicour_12}. The general spherically symmetric line element expressed in Bondi coordinates is:
\begin{equation}
ds^2=-\frac{V}{r}\mathrm{e}^{2\beta}du^2 - 2 \mathrm{e}^{2\beta} du dr + r^2(d \theta^2 + \sin^2 \theta d\varphi^2), \label{eq1}
\end{equation}

\noindent where $u=\mathrm{constant}$ denotes the null hypersurfaces that foliate the spacetime, and the metric functions $V,\beta$ depend on the coordinates $u,r$. The matter content is a scalar field $\phi=\phi(u,r)$ with the following energy momentum tensor: 
\begin{equation}
T_{\mu\nu}=\phi_{,\mu}\phi_{,\nu}-\frac{1}{2}g_{\mu\nu}\left[(\partial \phi)^2 + 2 U(\phi)\right], \label{eq2}
\end{equation}

\noindent where $(\partial \phi)^2=g^{\alpha\beta}\phi_{,\alpha}\phi_{,\beta}$ and $U(\phi)$ is the scalar field potential. 

We can recast the field equations in the following form: 
\begin{eqnarray}
%\begin{split}
& &\displaystyle \beta_{,r} = \frac{r}{4}\,\phi^2_{,r} \label{eq3}\\ 
\nonumber\\
& &\displaystyle V_{,r} = \mathrm{e}^{2\beta}\left(1-r^2 U(\phi)\right) \label{eq4} \\
\nonumber \\
& &\displaystyle (r\phi)_{,ur} - \frac{1}{2r}\left(r V \phi_{,r}\right)_{,r} + \frac{r}{2}\mathrm{e}^{2\beta}\frac{dU}{d\phi} = 0. \label{eq5}
%\end{split}
\end{eqnarray} 

\noindent We have adopted the unit system such that $8\pi G=c=1$. The first two equations are hypersurface equations, while the last is the Klein-Gordon equation. For the integration of Eqs. (\ref{eq3}) - (\ref{eq5}), we adopt the auxiliary field $\Phi$ defined by 
\begin{equation}
\Phi \equiv r\phi. \label{eq6}
\end{equation} 

\noindent The evolution scheme follows the typical hierarchical structure of the characteristic formulation. It starts with the initial data $\phi(u_0,r)$ from which after integrating Eq. (\ref{eq3}), we can determine $\beta(u_0,r)$. In the sequence, we can integrate Eq. (\ref{eq4}) to obtain $V(u_0,r)$. Next, Eq. (\ref{eq5}) determines $\phi_{,u}$ evaluated at $u=u_0$, allowing as a consequence to fix the scalar field at the next null hypersurface, or $\phi(u_0+\delta u,r)$. The whole cycle repeats, providing the evolution of spacetime.

For the sake of completeness, we present the main physical aspects associtated with the self-gravitating scalar field. We begin with the mass function, $m(u,r)$, that measures the total amount of mass-energy inside a sphere of radius $r$ at the hypersurface $u$. It is defined by
\begin{equation}
1-\frac{2 m}{r} = g^{\mu\nu} r_{,\mu} r_{,\nu} = \frac{V \mathrm{e}^{-2\beta}}{r}. \label{eq7}
\end{equation}

\noindent The formation of an apparent horizon signalizes the formation of a black hole. It occurs when $\beta \rightarrow \infty$, in another words when the expansion of outgoing null rays vanishes.

The Bondi mass, $M_B(u)$ is given by
\begin{equation}
M_B(u)=\lim_{r \rightarrow \infty} m(u,r). \label{eq8}
\end{equation}

\noindent Contrary to the ADM mass \cite{ADM}, the Bondi mass is not conserved but satisfies the Bondi formula \cite{bondi,gomez_jmp}
\begin{equation}
\frac{dM_B}{du} = -\frac{1}{2} \mathrm{e}^ {-2H(u)} N(u)^2, \label{eq9}
\end{equation} 

\noindent where $N(u)=\Phi_0(u)_{,u}$ is the news function, $\Phi_0(u)=\lim_{r \rightarrow \infty}\,\Phi$ and $H(u)=\lim_{r \rightarrow \infty}\,\beta$.

We summarize below the conditions near $r=0$ and near the future null infinity \cite{gomez_jmp}. Near the origin:
\begin{eqnarray}
\Phi(u,r)&=&\Phi_1(u)r+\Phi_2(u)r^2 + \mathcal{O}(r^3) \label{eq10}\\
\nonumber \\
\beta(u,r)&=&\mathcal{O}(r^2) \label{eq11}\\
\nonumber \\
V(u,r)&=&r + \mathcal{O}(r^3). \label{eq12}
\end{eqnarray}

\noindent Near the future null infinity:
\begin{eqnarray}
\Phi(u,r)&=& \Phi_0(u)+\frac{\Phi_{-1}(u)}{r}+\mathcal{O}(r^{-2}) \label{eq13}\\
\nonumber \\
\beta(u,r)&=&H(u)+\frac{\beta_{-2}(u)}{r^2}+\mathcal{O}(r^{-3}) \label{eq14}\\
\nonumber \\
V(u,r)&=&r\mathrm{e}^{2H}- 2\mathrm{e}^{2H}M_B(u)+\mathcal{O}(r^{-1}). \label{eq15}	 
\end{eqnarray}

\noindent The Bondi mass can be determined directly from the asymptotic expansion of the metric function $V(u,r)$ shown by Eq. (\ref{eq15}), or we can extend the integral form given by \cite{gomez_jmp} to include the potential
\begin{eqnarray}
M_B(u) = \frac{1}{4}\int_0^\infty\,\left[r V \mathrm{e}^{-2\beta}\left(\frac{\Phi}{r}\right)_{,r}^2 + 2 r^2 U(\phi) \right]dr. \nonumber \\
\label{eq16}
\end{eqnarray}   

\noindent We point out that the Bondi mass calculation is performed more accurately using the above integral expression.

%%%%%%%%%%%%%%%%%%%%%%%%%%%%%%%%%%%%%%%%%%%%%%%%%%%%%%%%%%%%
\section{The multidomain Galerkin-Collocation method}
%%%%%%%%%%%%%%%%%%%%%%%%%%%%%%%%%%%%%%%%%%%%%%%%%%%%%%%%%%%%

We describe here the numerical procedure based on the Galerkin-Collocation method in multiple nonoverlapping subdomains applied to the problem of a self-gravitating scalar field governed by Eqs. (\ref{eq3}) - (\ref{eq5}). We have divided the physical domain $\mathcal{D}: 0\leq r < \infty$ into $n$ subdomains 

\begin{eqnarray}
&&\mathcal{D}_1: r^{(0)}\leq r \leq r^{(1)};\;\;\mathcal{D}_2: r^{(1)} 
\leq r \leq r^{(2)},...,\nonumber \\
\nonumber \\
&&\mathcal{D}_n: r^{(n-1)} \leq r <\infty \nonumber
\end{eqnarray}

\noindent with $r^{(0)}=0$, and $r^{(1)},r^{(2)},..,r^{(n-1)}$ represent the interface between contiguous subdomains. We assume that $n \geq 2$.

\begin{figure*}[htb]
\includegraphics[scale=0.4]{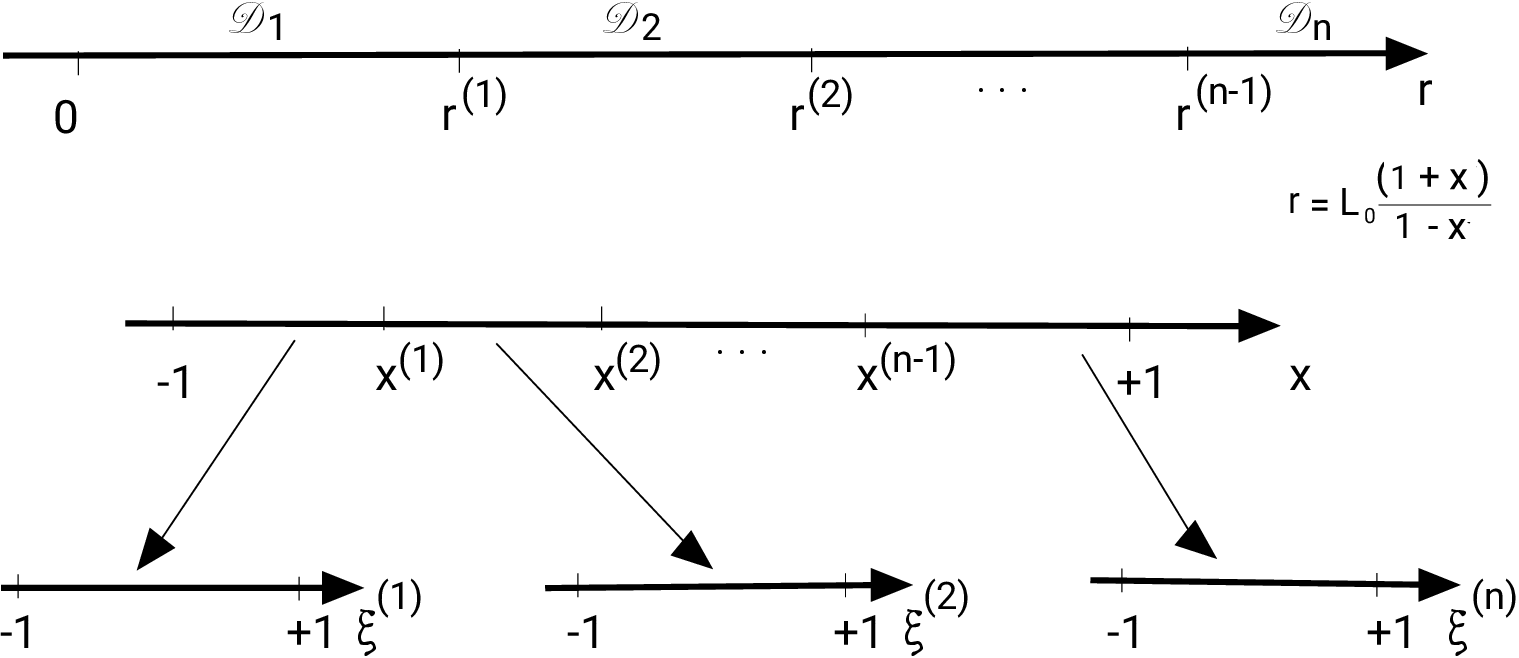}
	\caption{Basic scheme showing the subdomains $\mathcal{D}_1,\mathcal{D}_2,..,\mathcal{D}_n$ covered by $-1 \leq \xi_{j} \leq 1$, $k=1,2,..,n$, and the intermediate computational subdomain $-1 \leq x \leq 1$.}
\end{figure*}
To begin with the presentation of the multidomain Galerkin-Collocation procedure, we first establish the spectral approximations of the relevant functions in each subdomain and the appropriate transmission conditions. Let us consider the auxiliary scalar field $\Phi(u,r)$ (cf. Eq. (\ref{eq6})). At each subdomain, we have
\begin{eqnarray}
\Phi^{(l)}(u,r) = \sum^{N_l}_{k=0}a^{(l)}_k(u)\psi^{(l)}_k(r), \label{eq17}
\end{eqnarray}

\noindent where $l=1,2,..,n$ indicates the subdomain; $a^{(l)}_k(u)$ and $\psi^{(l)}_k(r)$, are the modes and the basis functions belonging to the $l$-th subdomain, respectively, and $N_l,\,l=1,2,..,n$ are the truncation orders that dictate the number of modes in each subdomain. 

According with the Galerkin method, the basis functions, $\psi^{(1)}_k(r)$ and $\psi^{(n)}_k(r)$, must satisfy the boundary conditions (\ref{eq10}) and (\ref{eq13}), respectively. For the the $n-2$ interior subdomains, the basis functions, $\psi_k^{(l)},\,l=2,3,..,n-1$, are the rational Chebyshev polynomials \cite{boyd} defined at each subdomain as we are going to see in the sequence. The first subdomain encompasses the condition (10), and we define the basis function as a linear combination of the rational Chebychev polynomials. For the last subdomain, the asymptotic condition (13) dictates that basis functions are the rational Chebyshev polynomials.

%To define the rational Chebyshev polynomials in each subdomain, we need to refer to Fig. 1. We show the basic scheme with the connection between the physical domain with the computational subdomains. First, the physical domain $\mathcal{D}: 0 \leq r < \infty$ is mapped into the interval $-1 \leq x \leq 1$ using the algebraic map \cite{boyd}
We define the rational Chebyshev polynomials in each subdomain converting the standard Chebyshev polynomials into rational functions in the radial coordinate $r$ using the mappings connecting the physical and the computational subdomains (see Fig. 1). First, the physical domain $\mathcal{D}: 0 \leq r < \infty$ is mapped into the interval $-1 \leq x \leq 1$ using the algebraic map \cite{boyd}
\begin{equation}
r = L_0 \frac{(1+x)}{1-x}, \label{eq18}
\end{equation}

\noindent where $L_0$ is the map parameter and $x^{(1)},x^{(2)},..,x^{(n-1)}$ are the interface between contiguous subdomains as shown in Fig. 1. Second, we introduce linear transformations to define the computational subdomains parameterized by $-1 \leq \xi^{(l)} \leq 1$, $l=1,2,..,n$:
\begin{eqnarray}
x(\xi^{(l)})=\frac{1}{2}\,\left[\left(x^{(l)}-x^{(l-1)}\right)\xi^{(l)}+x^{(l)}+x^{(l-1)}\right], \label{eq19}
\end{eqnarray} 

\noindent with $l=1,2...,n$ where $x^{(0)}=-1$ and $x^{(n)}=1$. These computational subdomains are the loci of the collocation points that are mapped back to $r_k$ in the physical domain $\mathcal{D}$.

The rational Chebyshev polynomials are defined in each subdomains as
\begin{eqnarray}
TL_k^{(l)} = T_k\left(\xi^{(l)}=\frac{\mathrm{a}^{(l)}r+\mathrm{b}^{(l)}}{(r+L_0)}\right), \label{eq20}
\end{eqnarray}

\noindent where 
\begin{eqnarray}
\mathrm{a}^{(l)}&=&\frac{2L_0+r^{(l)}+r^{(l-1)}}{r^{(l)}-r^{(l-1)}}, \label{eq21}\\
\nonumber \\ 
\mathrm{b}^{(l)}&=&-\frac{2r^{(l)}r^{{(l-1)}}+L_0\left(r^{(l)}+r^{(l-1)}\right)}{r^{(l)}-r^{(l-1)}}. \label{eq22}
\end{eqnarray}

\noindent $T_k(\xi)$ is the Chebyshev polynomial of kth order, $r^{(l-1)} \leq r \leq r^{(l)}$ corresponds to $-1 \leq \xi^{(l)} \leq 1$ for all $l=1,2,..,n$ with $r^{(0)}=0$ and $r^{(n)}$ is located at the infinity.

We can now define the basis functions for the spectral approximation (\ref{eq17}). As we have mentioned, the rational Chebyshev polynomials (\ref{eq20}) are the basis for the subdomains $\mathcal{D}_l, l=2,..,n$. The basis functions defined in the first subdomains $\mathcal{D}_1$ are given by
\begin{eqnarray}
\psi_k^{(1)}(r) &=& \frac{1}{2}\left(TL_{k+1}^{(1)}(r) + TL_{k}^{(1)}(r)\right). \label{eq23}
%\nonumber \\
%\psi_k^{(n)}(r) &=& TL_{k}^{(n)}(r). \label{eq24}
\end{eqnarray}

\noindent From the above definition, we have $\psi_k^{(1)}(r)=\mathcal{O}(r)$ that satisfies the condition (\ref{eq10}).  

The next step is to establish the transmission or matching conditions to guarantee that all pieces of Eq. (\ref{eq17}) represent the same function $\Phi(u,r)$ but viewed in each subdomain. We adopt the patching method \cite{canuto_88} demanding that a function and all their $(d-1)$th spatial derivatives must be continuous at the interface of contiguous subdomains, where $d$ is the highest derivative order. In the present case, all spatial derivatives involve only the radial coordinate $r$, and since for the field $\Phi(u,r)$ we have $d=2$, the transmission conditions are
\begin{eqnarray}
\Phi^{(l)}\left(u,r^{(l)}\right)&=&\Phi^{(l+1)}\left(u,r^{(l)}\right) \label{eq24}\\
\nonumber \\
\left(\frac{\partial \Phi^{(l)}}{\partial r}\right)_{r^{(l)}}&=&\left(\frac{\partial \Phi^{(l+1)}}{\partial r}\right)_{r^{(l)}} \label{eq25}
\end{eqnarray}

\noindent with $l=1,2,..,n$. We remark that the number of transmission conditions, together with the number of collocation points, must be equal to the total number of modes $a_k^{(l)}$.

The spectral approximations for the functions $\beta(u,r)$ and $V(u,r)$ are
\begin{eqnarray}
\beta^{(l)}(u,r) &=&  \sum^{\bar{N}_l}_{k=0}b^{(l)}_k(u)\chi^{(l)}_k(r) \label{eq27} \\
\nonumber \\
V^{(l)}(u,r) &=&  r + \sum^{\bar{N}_l}_{k=0}c^{(l)}_k(u)r\tilde{\chi}^{(l)}_k(r), \label{eq28}
\end{eqnarray}

\noindent where $b^{(l)}_k(u)$ and $c^{(l)}_k(u)$ are the unknown modes associated with the metric functions $\beta(u,r)$ and $V(u,r)$ defined in each subdomain. Notice that the truncation orders $\bar{N}_l$ can be distinct from those of the auxiliar field approximation meaning that we have, in general two sets of collocation points. The basis functions $\chi^{(1)}_k(r), \tilde{\chi}^{(1)}_k(r)$ and $\chi^{(n)}_k(r),\tilde{\chi}^{(n)}_k(r)$ satisfy the conditions (\ref{eq11}), (\ref{eq12}) and (\ref{eq14}), (\ref{eq15}), respectively (see Appendix). And the transmission conditions are:
\begin{eqnarray}
\beta^{(l)}\left(u,r^{(l)}\right)&=&\beta^{(l+1)}\left(u,r^{(l)}\right) \label{eq29} \\
\nonumber \\
V^{(l)}\left(u,r^{(l)}\right)&=&V^{(l+1)}\left(u,r^{(l)}\right), \label{eq30}
\end{eqnarray}

\noindent with $l=1,2,..,n-1$.

The goal of any spectral method is to approximate a set of evolution partial differential equations into a finite set of ordinary differential equations. Considering the Eq. (\ref{eq5}) the result is a dynamical system for the modes $a_k^{(l)}(u)$, $l=1,2,..,n$. On the other hand,  we approximate the hypersurface equations (\ref{eq3}) and (\ref{eq4}) to sets of algebraic equations for the all modes $b_k^{(l)}(u)$ and $c_k^{(l)}(u)$. In what follows, we have generalized the procedure outlined in Refs. \cite{barreto_19,crespo_19}.

For the sake of simplicity, we consider the hypersurface equation (\ref{eq3}). After substituting the spectral approximation (\ref{eq17}) and (\ref{eq27}), we obtain the residual equations belonging to each subdomain. Esquematically, we have 
\begin{eqnarray}
\mathrm{Res}^{(l)}_\beta(u,r) = \sum_{k=0}^{\bar{N}_l}\,b_k^{(l)}(u) \chi_k^{(l)}(r) - \frac{r}{2}\,\left(\frac{\Phi^{(l)}}{r}\right)_{,r}^2, \label{eq31}
\end{eqnarray}

\noindent with $l=1,2,..,n$. We follow the prescription of the Collocation method to determine a set of algebraic equations for the modes $b_k^{(l)}(u)$, meaning to vanish the residual equations represented by  (\ref{eq29}) at the corresponding collocation points located in each subdomain. Then   
\begin{eqnarray}
\mathrm{Res}^{(l)}_\beta(u,r^{(l)}_j) = 0. \label{eq32}
\end{eqnarray}

\noindent where $r^{(l)}_j$ are the collocation points in the physical domain related to the collocation points $\xi_j^{(l)}$ in the computational subdmains by
\begin{eqnarray}
r^{(l)}_j=\frac{\mathrm{b}^{(l)}-L_0\xi_j^{(l)}}{\xi_j^{(l)}-\mathrm{a}^{(l)}}, \label{eq33}
\end{eqnarray}

\noindent with $\mathrm{a}^{(l)},\mathrm{b}^{(l)}$ given by the relations (\ref{eq21}) and (\ref{eq22}), respectively. The total amount of the collocation points, $\xi^{(l)}_j$ in the computational subdomains must be  

\[
\underbrace{\sum_{l=1}^n \bar{N}_l+n}_{\substack{\text{number of}\\\text{modes $b_k^{(l)}$}}} - \underbrace{(n-1)}_{\substack{\text{transmission}\\\text{conditions}}} = 
\underbrace{\sum_{l=1}^n \bar{N}_l+1}_{\substack{\text{number of collocation}\\\text{points}}}.
\]

\noindent Therefore, we have $\sum_{l=1}^n \bar{N}_l+1$ equations (\ref{eq32}). These equations together with the transmission conditions (\ref{eq29}) form a closed algebraic system for the modes $b_k^{(l)}$ whose solution can be read as   
\begin{eqnarray}
%b_k^{(l)}=F_k^{(l)}(dr\Phi^{(q)}_j), \label{eq36}
b_k^{(l)}=F_k^{(l)}\left(\left[\left(\frac{\Phi}{r}\right)_{,r}\right]_j^{(q)}\right), \label{eq36}
\end{eqnarray}

\noindent where $k=0,1,..,\bar{N}_l$, $l,q=1,2,..,n$ and $\left[\left(\Phi/r\right)_{,r}\right]_j^{(q)}$ are the values of $\left(\Phi/r\right)_{,r}$ at the collocation points of the $q$th-subdomain. It is possible that $q \neq l$ meaning that the modes belonging to the $l$th subdomain can depend on the values of $\left(\Phi/r\right)_{,r}$ evaluated at other subdomains than the $l$th.

We choose $\xi_j^{(l)}$ as the Chebyshev-Gauss-Lobatto points \cite{peyret} in each subdomain. The distribution of the $\sum_{l=1}^n \bar{N}_l+1$ collocation points in the subdomains is indicated in Table 1. Accordingly, in the first subdomain, we have
\begin{eqnarray}
\xi_j^{(1)} &=& \cos\left(\frac{j \pi}{\bar{N}_l+2}\right),\;\;j=1,2,..,\bar{N}_l+1, \label{eq34}
\end{eqnarray}

\noindent and in the remaining subdomains
\begin{eqnarray}
\xi_j^{(l)} &=& \cos\left(\frac{j \pi}{\bar{N}_l+1}\right),\;\;j=1,2,..,\bar{N}_l,\label{eq35}
\end{eqnarray} 

\noindent for all $l=2,3,..,n$.  Notice that for the above set of collocation points, we have placed the interfaces $r^{(1)},r^{(2)},..,r^{(n-1)}$ as belonging to the corresponding subsequent subdomains, $\mathcal{D}_2,\mathcal{D}_3,..,\mathcal{D}_n$, respectively

\begin{table}
	\centerline{Table 1}
	\medskip
	\begin{center}
	\centering
	\begin{tabular}[c]{ l|c|c|c|c|r }
		\hline
		  %\\
		&$\mathcal{D}_1$ & $\mathcal{D}_2$ &  \hspace{0.5cm} ... \hspace{0.5cm}  & $\mathcal{D}_{n-1}$ &$\mathcal{D}_{n}$ \\
		  %\\
		\hline
		\hline
		   %\\
		$\beta $ & $\bar{N}_1+1$ & $\bar{N}_2$ & ... & $\bar{N}_{n-1}$ & $\bar{N}_n$ \\
		%\vspace{0.2cm}
		%\hline
		$V$ & $\bar{N}_1+1$ & $\bar{N}_2$ & ... & $\bar{N}_{n-1}$ & $\bar{N}_n$\\
		%   \\
		%\hline
		%\\
		$\Phi$ & $N_1$ & $N_2-1$ & ... & $N_{n-1}-1$ & $N_n$ \\
		\hline
	\end{tabular}
\caption{Distribution of the collocation points in each subdomains according to corresponding truncation orders.}
\end{center}
\end{table}

For the metric function $V(u,r)$, we perform the same procedure outlined previously, taking into account the collocation points (\ref{eq34}) and (\ref{eq35}). We obtain an algebraic system for the modes $c_k^{(l)}$ whose solution can be written in the following way 
\begin{eqnarray}
c_k^{(l)}=G^{(l)}_k(\beta^{(q)}_j,\Phi^{(q)}_j), \label{eq37}
\end{eqnarray}

\noindent where $k=0,1,.,\bar{N}_l$ and $l,q=1,2,..,n$. Here $\beta^{(q)}_j$ and $\Phi^{(q)}_j$ are the values of $\beta$ and $\Phi$ at the collocation points in the $q$th subdomains.   

The final step is to obtain the dynamical system that approximates the Klein-Gordon equation (\ref{eq5}). We have followed the same procedure described previously first by establishing the corresponding residual equations in each subdomain. We have,
\begin{eqnarray}
\mathrm{Res}^{(l)}(u,r)=&&\sum_{k=0}\frac{da_k^{(l)}}{du}\psi_k^{(l)}(r) - \frac{1}{2r}\left[r V \left(\frac{\Phi}{r}\right)_{,r}\right]^{(l)}  \nonumber \\
\nonumber \\
&&+\frac{r^2}{2}\mathrm{e}^{2\beta^{(l)}}\left(\frac{dU}{d\Phi}\right)^{(l)}, \label{eq38}
\end{eqnarray}

\noindent where $l=1,2..,n$, the functions $\Phi,V,\beta$ above are given by their respective spectral approximations in each subdomain. Second, we proceed by vanishing the residual equations (\ref{eq38}) at the collocation points defined in each subdomain and taking into account the transmission conditions (\ref{eq24}) and (\ref{eq25}). From these equations, we obtain the following set of equations:
\begin{equation}
\frac{da_k^{(l)}}{du}=L_k^{(l)}(\beta_j^{(q)},V_j^{(q)},\Phi_j^{(q)},..), \label{eq39}
\end{equation}

\noindent where $k=0,1,..,N_l$ and $l,q=1,2,..,n$. The RHS depends on the values of $\beta,V,\Phi,..$ at the collocation points not only in the $l$th subdomain.

For the residual equation (\ref{eq38}), we have considered another set of collocation points.  The number of this new set of collocation is connected with the transmission conditions according to 

\[
\underbrace{\sum_{l=1}^n N_l+n}_{\substack{\text{number of}\\\text{modes $a_k^{(l)}$}}} - \underbrace{2(n-1)}_{\substack{\text{transmission}\\\text{conditions}}} = \underbrace{\sum_{l=1}^n N_l-n+2}_{\substack{\text{number of collocation}\\\text{points}}},
\]

\noindent where $n \geq 2$. We have chosen the Chebychev-Gauss-Lobatto points and presented their distribution in each subdomain in Table 1. Then, for the first, $l=1$ and the last, $l=n$ subdomains, we have $N_l$ collocation points: 

\begin{eqnarray}
\xi_j^{(l)} = \cos\left(\frac{j \pi}{N_l}\right),\;\; \text{if $l = 1,n$}, \label{eq40}
\end{eqnarray}

\noindent where $j=0,1,..,N_1$ and $j=1,2,..,N_n-1$, for the first and the last sudodmains, respectively. And for all other subdomains, it follows  
\begin{eqnarray}
\xi_j^{(l)} &=& \cos\left(\frac{j \pi}{N^{(l)}-1}\right),\;\; \text{if $l = 2,..,n-1$}, \label{eq41}
\end{eqnarray} 

\noindent where $j=1,2..,N_l-1$. Again, it is worth of mentioning that we have placed the interfaces $r^{(1)},r^{(2)},..,r^{(n-1)}$ as belonging to the corresponding subsequent subdomains, $\mathcal{D}_2,\mathcal{D}_3,..,\mathcal{D}_n$, respectively. We remark that by choosing distinct truncation orders, $N_l$ and $\bar{N}_l$ means two separate grids in each subdomain.

At this point, it is necessary to remark about possible additional interface conditions associated to the evolution equation (\ref{eq5}). Following Kopriva \cite{kopriva_86}, Canuto et al. \cite{canuto_88}, domain decomposition in hyperbolic problems require special care with additional conditions at the interfaces. In the present case, we take advantage of placing the interface points belonging to the collocation points of the subsequent subdomain to assume that
\begin{eqnarray}
\left(\Phi\right)^{(l+1)}_{ur}-\frac{1}{2r^{(l)}}\left[r V \left(\frac{\Phi}{r}\right)_{,r}\right]_{,r}^{(l+1)} +\frac{r^{(l)}}{2} \times\nonumber  \\
\nonumber \\ \mathrm{e}^{2\beta^{(l+1)}}\left(\frac{dU}{d\Phi}\right)^{(l+1)}=0, \label{eq42}
\end{eqnarray} 

\noindent where $l=1,2,..,n-1$. The above conditions are equivalent to the \textit{upwind} scheme \cite{kopriva_86,kopriva_89}. Another possibility is to adopt the average procedure that consists of collecting both contiguous subdomains contributions. For instance, at the interface $r^{(l)}$ we would have
%
%\begin{widetext}
\begin{eqnarray}
\left(\Phi\right)^{(l+1)}_{ur}-\frac{1}{4r^{(l)}}\Bigg\{\left[r V \left(\frac{\Phi}{r}\right)_{,r}\right]_{,r}^{(l+1)} + \left[r V \left(\frac{\Phi}{r}\right)_{,r}\right]_{,r}^{(l)}\Bigg\} + \nonumber \\
\frac{r^{(l)}}{4}\Bigg\{\mathrm{e}^{2\beta^{(l+1)}}\left(\frac{dU}{d\Phi}\right)^{(l+1)} + \mathrm{e}^{2\beta^{(l)}}\left(\frac{dU}{d\Phi}\right)^{(l)}\Bigg\}=0, \label{eq43}
\end{eqnarray} 
%\end{widetext}

\noindent We have implemented the above average condition in Ref. \cite{crespo_19} for a code with two subdomains to integrate the field equations in the affine characteristic formulation. Although the average interface condition provided a stable code, its implementation in the realm of multidomain decomposition requires careful attention. For this reason, we have adopted the upwind interface condition (\ref{eq42}) not only due to its simplicity but also for generating a stable and accurate code.

\section{Numerical results: code validation and applications}

We divided the present Section into two parts. In the first part, we deal with two numerical tests: the verification of the Newman-Penrose constant and the Bondi formula for the energy conservation. Both tests are necessary to validate the multidomain GC algorithm. In particular, we are interested in the influence of increasing the number of subdomains and the number of collocation points on the code convergence. In the second part, we applied the multidomain algorithm to recast the black hole scaling law connected with the critical gravitational collapse originally discovered by Choptuik \cite{choptuik}.

In the numerical tests to validate the code, we have integrated the dynamical system with a standard fourth-order Runge-Kutta method, whereas to investigate the critical collapse, we have used the Cash-Karp \cite{cash-karp} adaptive integrator. For most cases, the truncation order $\bar{N}_l$ has an increase of about 30\% of the corresponding $N_l$.

\subsection{Validating the code}

In dealing with the multidomain algorithm, we have to select the number of subdomains, the interface locations, and the map parameter. We adopted the division of the intermediate computational domain $-1 \leq x \leq 1$ (cf. Fig. 1) in equal parts. For instance, if we have two subdomains, $x^{(1)}=0$ implying that $r^{(1)}=L_0$, in the case of three subdomains, $x^{(1)}=-1/3$ and $x^{(2)}=1/3$ implying in $r^{(1)}=L_0/2$ and $r^{(2)}=2 L_0$, respectively, and so on. Consequently, the information about the location of the interfaces is determined by the map parameter $L_0$. In Table 2, we indicate the ranges of each subdomain for the sake of clarity.

\begin{table}
	\centerline{Table 2}
	\medskip
	\begin{center}
		\centering
		\begin{tabular}[c]{c |c |c |c}
			\hline
			%\\
			$\mathcal{D}_1$ & $\mathcal{D}_2$ & $\mathcal{D}_3$ & $\mathcal{D}_{4}$ \\
			%\\
			\hline
			\hline
			%\\
			$0 \leq r \leq L_0$ & $L_0 \leq r < \infty$ & - & - \\
			%\\
			\hline
			%\\
			$0 \leq r \leq \frac{1}{2}L_0$ & $\frac{1}{2}L_0 \leq r \leq 2L_0$ & $2L_0 \leq r < \infty$ & - \\
			%\\
			\hline
			%\\
			$0 \leq r \leq \frac{1}{3}L_0$ & $\frac{1}{3}L_0 \leq r \leq L_0$ & $L_0 \leq r \leq 3L_0$ & $3L_0 \leq r < \infty$ \\
			%\\
			\hline
		\end{tabular}
	\caption{We show above the subdomain boundaries in terms of the map parameter after assuming that the intermediate computational domain is divided equally}
	\end{center}
\end{table}

The first numerical test is the verification of the Newman-Penrose constant \cite{NP}, $c_{NP}$. We obtain this quantity from the asymptotic expansion of the auxiliary scalar field given by Eq. (\ref{eq13}). The field equations dictate that
\begin{equation}
c_{NP} \equiv \Phi_{-1}(u) = \mathrm{constant}. \label{eq45}
\end{equation}

We proceed with the following the initial data
\begin{equation}
\Phi_0(r) = \frac{A_0 r}{k_1+(r-k_2)^2}, \label{eq46}
\end{equation}

\noindent where $k_1$ and $k_2$ are positive constants and $A_0$ is the scalar field initial amplitude. Taking into account the previous equations, we identify the NP constant as 
\begin{equation}
c_{NP} = A_0. \label{eq47}
\end{equation}

\noindent We compare this value with the approximate NP constant provided by the spectral approximation of the scalar field at the $n$th or the last subdomain (cf. Eq (\ref{eq17})). It means that we express the approximate $c_{NP}$ in terms of the modes $a_k^{(n)}(u)$ with
\begin{equation}
c_{NP}(u) = - \lim_{r \rightarrow \infty}\left(r^2\frac{\partial \Phi^{(n)}}{\partial r}\right). \label{eq48}
\end{equation}  

\noindent For the numerical experiments, we set $k_1=0.5, k_2=2.0$, and $A_0=0.27$ representing an initial pulse that, although dispersing, is close to the formation of a black hole ($A_0  \gtrsim 0.28$). We integrate the dynamical system (\ref{eq39}) along with the Eqs. (\ref{eq36}) and (\ref{eq37}) using a forth-order Runge-Kutta integrator with a stepsize of $10^{-4}$. We calculate the relative deviation $\delta c_{NP}$ according to 
\begin{eqnarray}
\delta c_{NP} = \frac{|c_{NP}(0)-c_{NP}(u)|}{c_{NP}(0)} \times 100, \label{eq49}
\end{eqnarray}

\noindent where $c_{NP}(0)$ is given by Eq. (\ref{eq47}). 

We show in Fig. 2 the log-linear plots of $\delta c_{NP}$ in terms of the total truncation order of the scalar field for two, three, and four subdomains for some values of $L_0$. Notice that for two subdomains, the convergence is not well defined for $L_0=0.5$ and $1$; for $L_0=2$,  the exponential decay of $\delta c_{NP}$ becomes evident. From Table 2, $r \geq L_0$ defines the second subdomain. In the case of three subdomains, we noticed the exponential decay of $\delta c_{NP}$ for all values of $L_0$. In particular, this decay is more effective for $L_0=1$ where $\delta c_{NP} \simeq 10^{-10} \%$ for total truncation order of $N=60+60+80=200$. By considering four subdomains, we observe that the exponential convergence for $L_0=0.5,1$ and $2$, but the last value produces a better decay.

The second numerical test consists in verifying the validity of the Bondi formula (9) after conveniently rewritten by introducing the function $C(u)$ \cite{gomez_jmp}:
\begin{eqnarray}
C(u) = \frac{1}{M_B(u_0)}\Bigg[M_B(u)-M_B(u_0) + \nonumber \\
\nonumber \\
+ \frac{1}{2}\,\int_{u_0}^u \mathrm{e}^{-2H(u)} N(u)^2\,du \Bigg] \times 100 \label{eq50}
\end{eqnarray}

\noindent $C(u)$ measures the deviation from the global energy conservation described by the Bondi formula with $C(u)=0$ for the exact solution. For this test, we have considered a Gaussian initial data
\begin{equation}
\Phi_0(r) = A_0 r \mathrm{e}^{-(r-r_0)^2/\sigma^2}. \label{eq51}
\end{equation}

\noindent Here, we have set $r_0=0.5$, $\sigma=1$ and $A_0=1.5$ which is close to evolutions that result in the  formation of an apparent horizon. 
We evolve the system as before using a fourth-order Runge-Kutta integrator but with stepsize $5.0\times 10^{-5}$. The integral on the RHS of Eq. (\ref{eq50}) is calculated with the Newton-Cotes formulae \cite{NC}.
In Fig. 3, we have fixed $L_0=0.5$ and display the decay of $C(u)$ for the case of two (blue boxes), three (red triangles), and four subdomains (black circles). In general, no matter is the division of subdomains, $C(u)$ decays exponentially, indicating a rapid convergence of the spectral code. For the present initial data and value of the map parameter, the division in three subdomains seems slightly better. However, for a global truncation order of $N \geq 200$, the error reaches the machine precision. 

An alternative way of rewriting the Bondi formula is to approximate the LHS of Eq. (9) that results in 
\begin{eqnarray}
\Delta M_B(u) &=& \Big|\frac{1}{2 \delta u}(M_B(u+\delta u)-M_B(u-\delta u)) \nonumber\\
\nonumber \\
&+& \frac{1}{2} \mathrm{e}^{-2H(u)} N(u)^2\Big|. \label{eq52}
\end{eqnarray}

\noindent Fig. 4 illustrates the rapid decay of the error, where the three and four subdomains codes display equivalent results. 

It is worth mentioning that the convergence depends on several factors: the location of the interfaces, the map parameter, the number of collocation points in each subdomain, and of course of the number of subdomains. As mentioned, we generate the results displayed in figures 2, 3, and 4 with the interfaces chosen such that the intermediate computational domain is divided equally. It is possible to make distinct choices for the interface locations aiming specific goals, for instance, to determine precisely the Bondi mass at the moment the apparent horizon forms.
\begin{figure}[h!]
\begin{center}
\includegraphics[width=8.5cm,height=5.5cm]{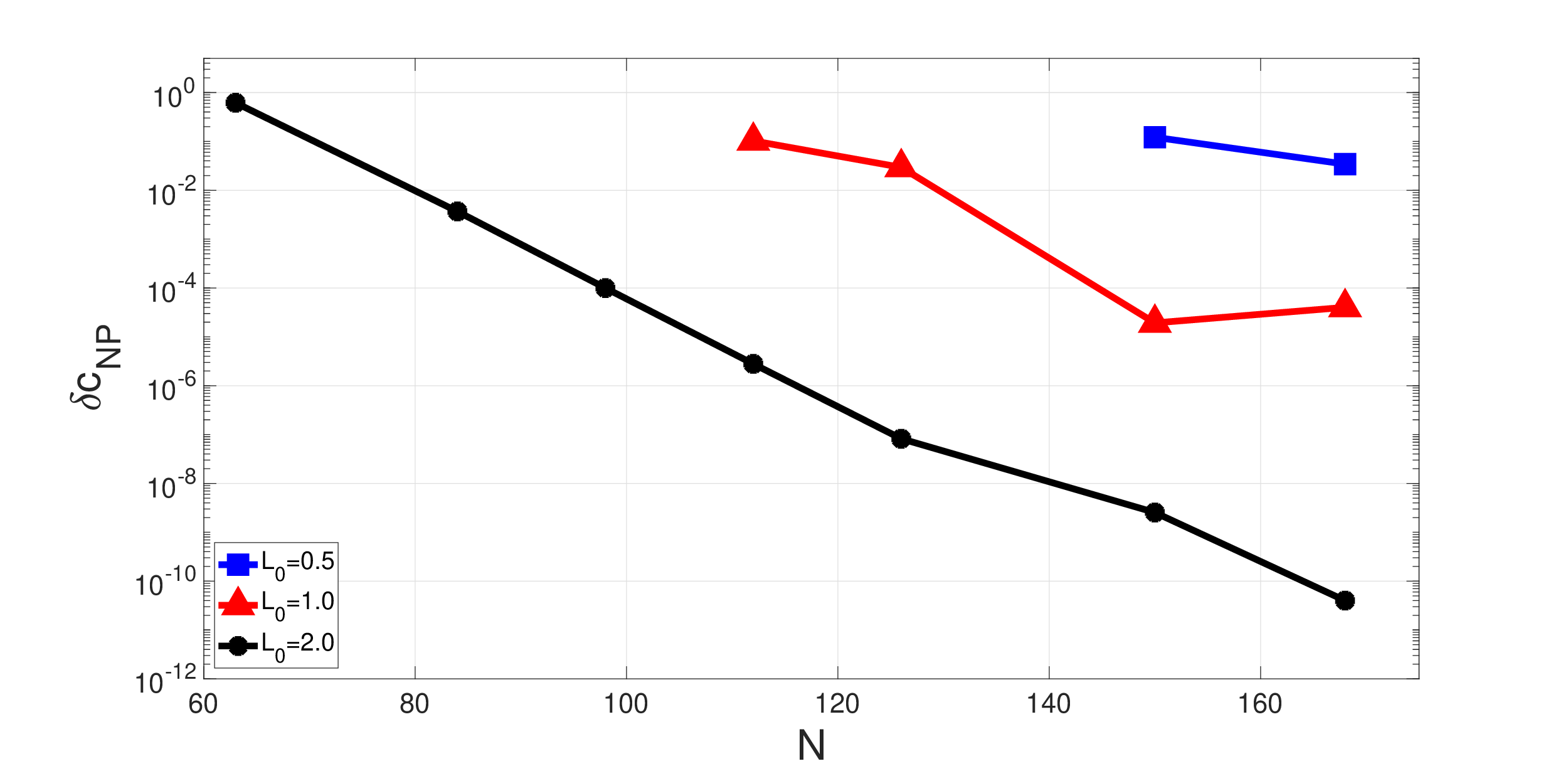}
\includegraphics[width=8.5cm,height=5.5cm]{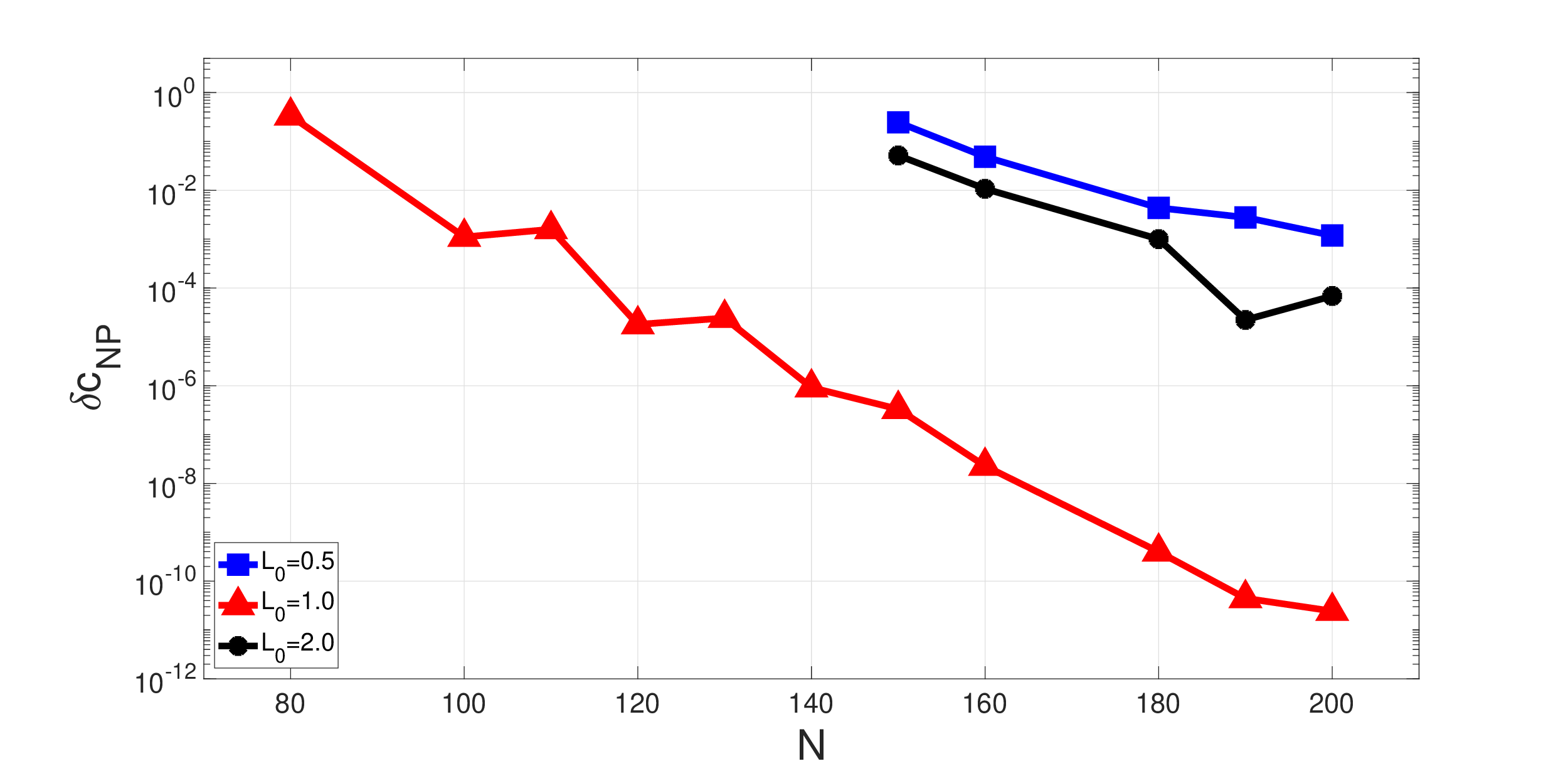}
\includegraphics[width=8.5cm,height=5.5cm]{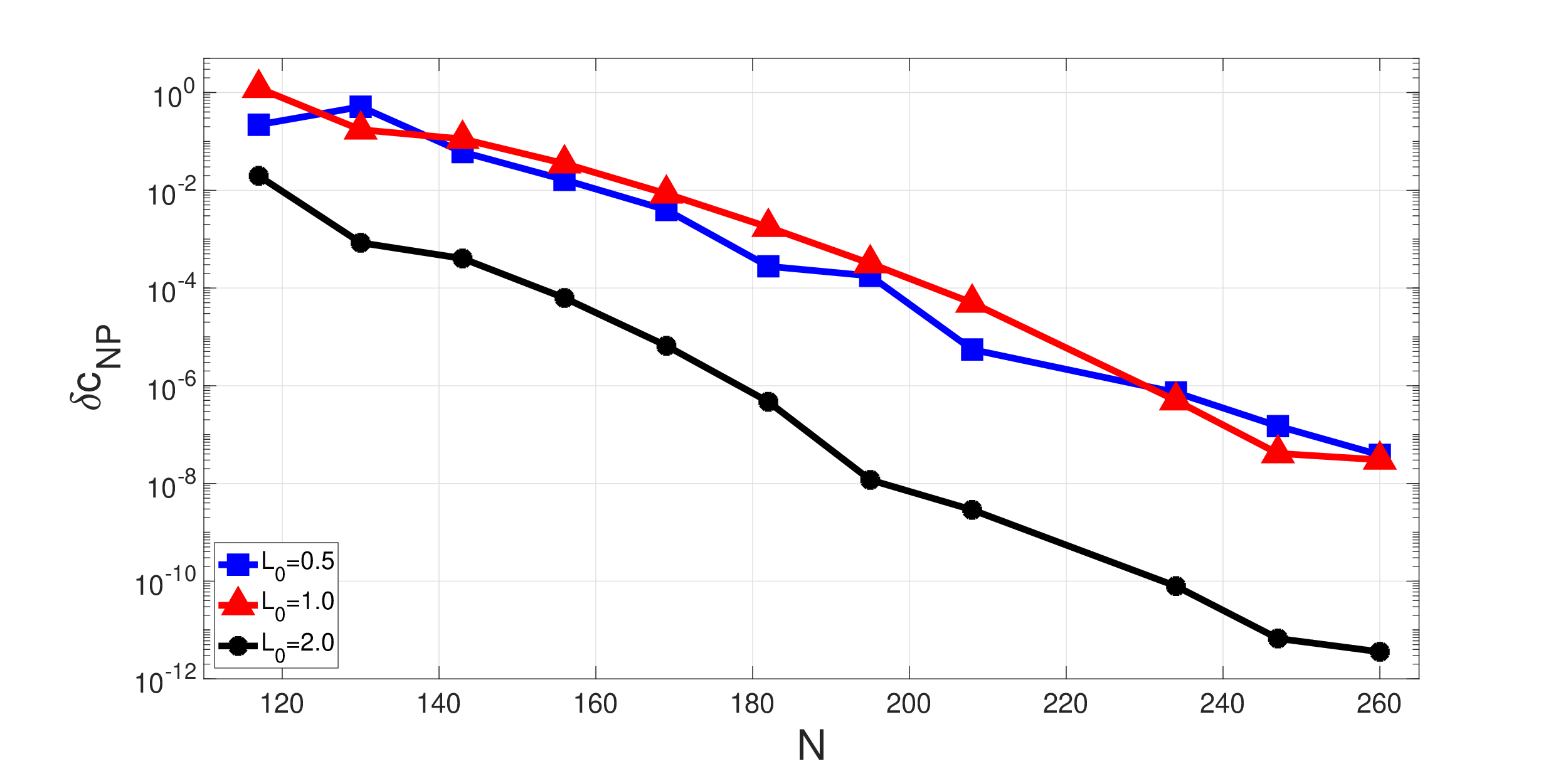}
\caption{We present from the upper to bottom panels the exponential decay of $(\delta c_{NP})_{max}$ for the cases of two, three, and four subdomains, respectively.  The graphs with boxes, triangles, and circles correspond to the map parameter $L_0=0.5,1.0$ and $2.0$, respectively. $N$ is the total number of collocation points.}
\end{center}
\end{figure}
\begin{figure}[htb]
	\includegraphics[width=9cm,height=6.5cm]{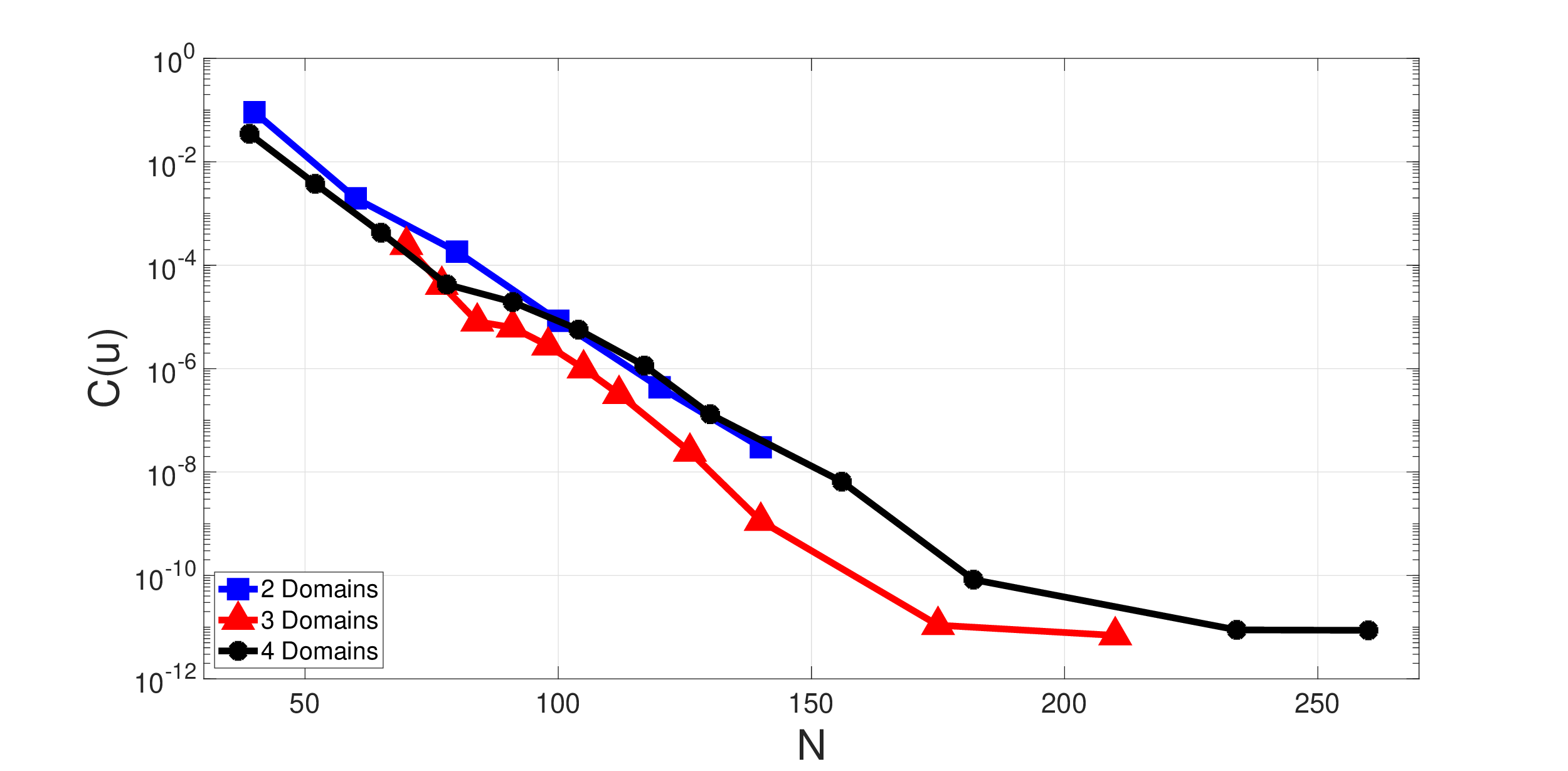}
	\caption{Exponential decay of $C_{\mathrm{max}}$ for the cases of two (boxes), three (triangles), and four subdomains (circles).  We have fixed the map parameter $L_0=0.5$. $N$ is the total number of collocation points.}
\end{figure}
\begin{figure}[htb]
	\includegraphics[width=9cm,height=6.5cm]{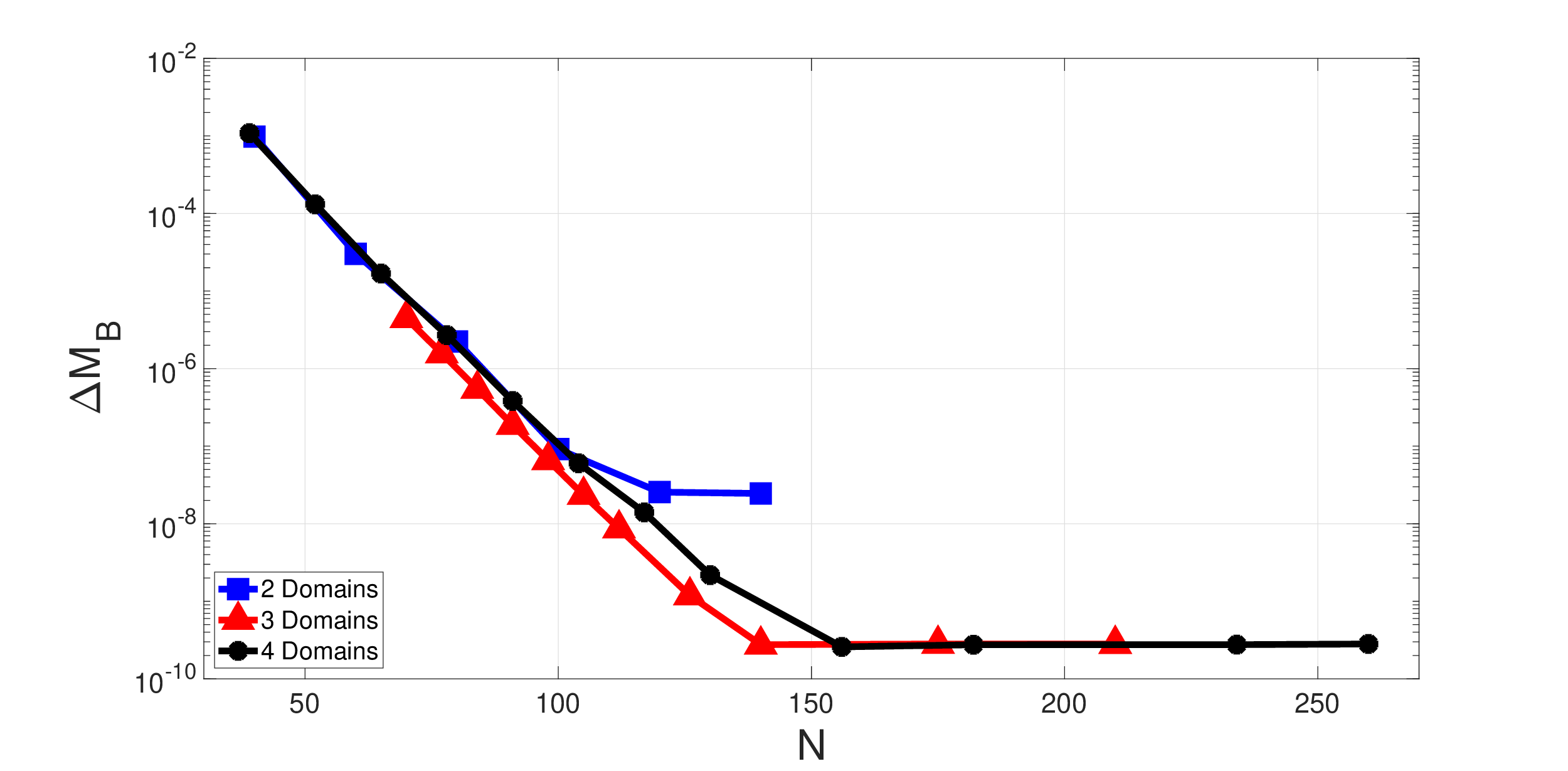}
	\caption{Exponential decay of $(\Delta M_B)_{\mathrm{max}}$ (cf. Eq. (\ref{eq52}) for the cases of two (boxes), three (triangles), and four subdomains (circles).  We have fixed the map parameter $L_0=0.5$. $N$ is the total number of collocation points.}
\end{figure}
\begin{figure}
\includegraphics[width=15.5cm,height=10.5cm]{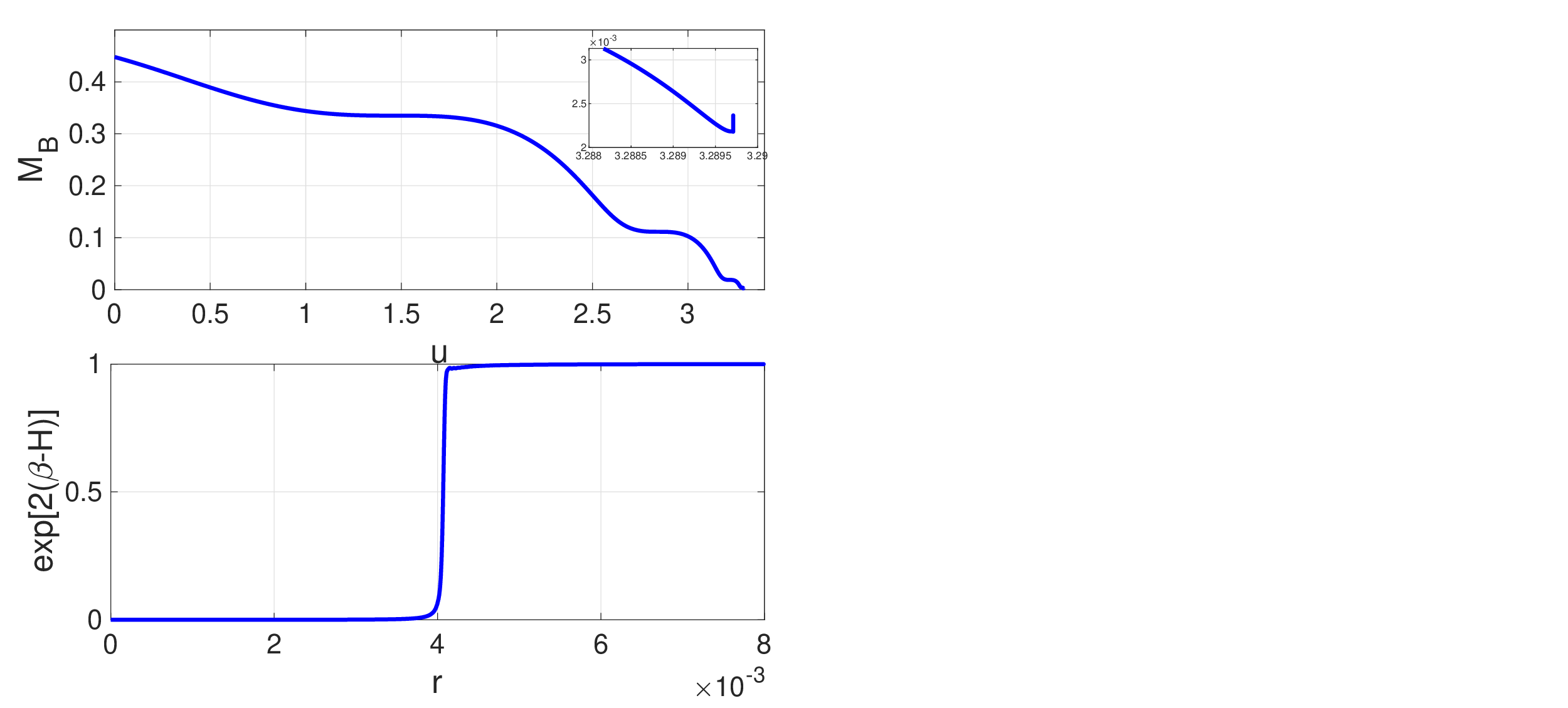}
\caption{In the upper panel, we illustrate the Bondi mass behavior for a supercritical solution until the apparent horizon forms. In the lower plot, we present the typical step function predicted by Christodoulou prior to the formation of the apparent horizon. In both cases, apparent horizon mass $m_{AH} = r_{AH}/2$ and the Bondi mass are approximately equal. Here $M_{BH} \approx 2.18 \times 10^{3}$ corresponding to $A_0=1.5129325$ and $L_0=0.1$.}
\end{figure}
\begin{figure*}[h!]
\includegraphics[width=5.5cm,height=4.5cm]{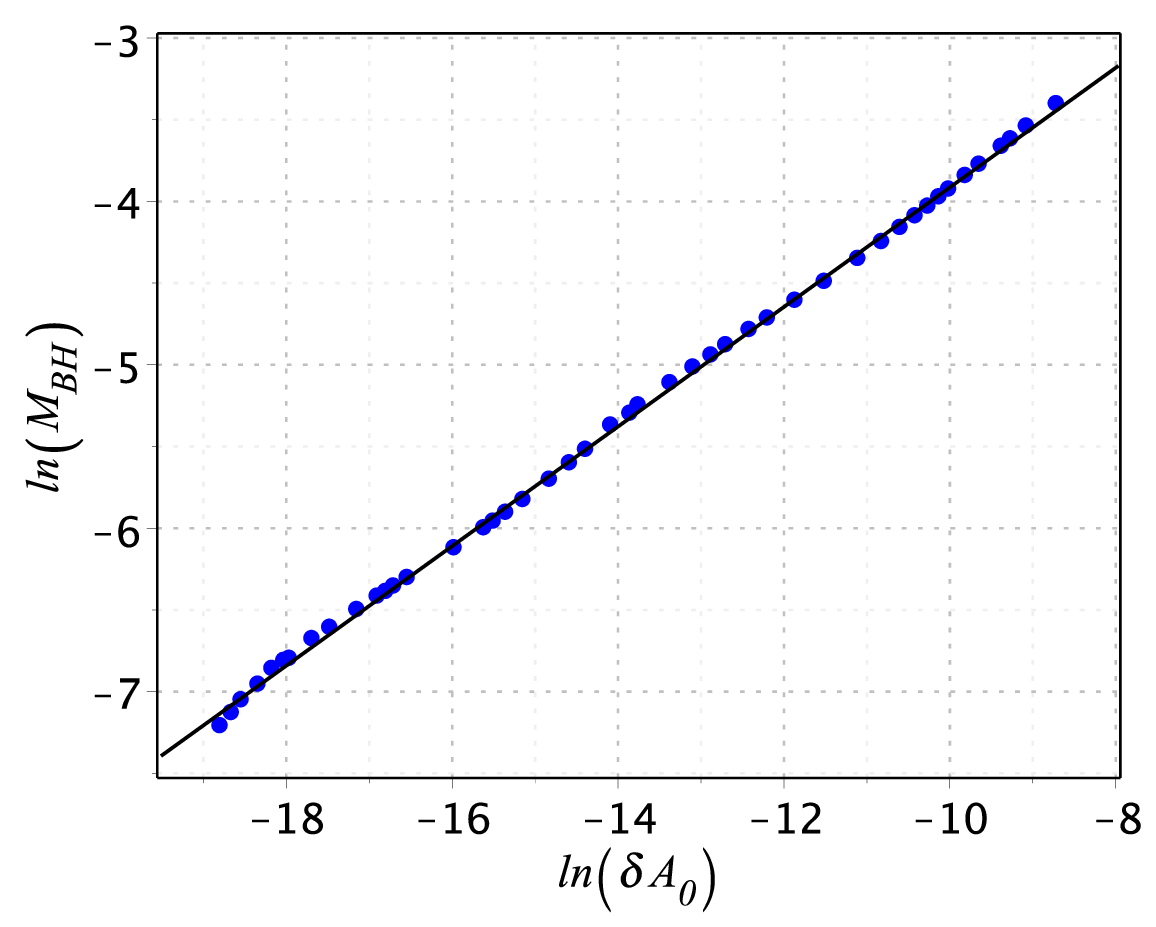}\hspace{0.5cm}
\includegraphics[width=6cm,height=4.45cm]{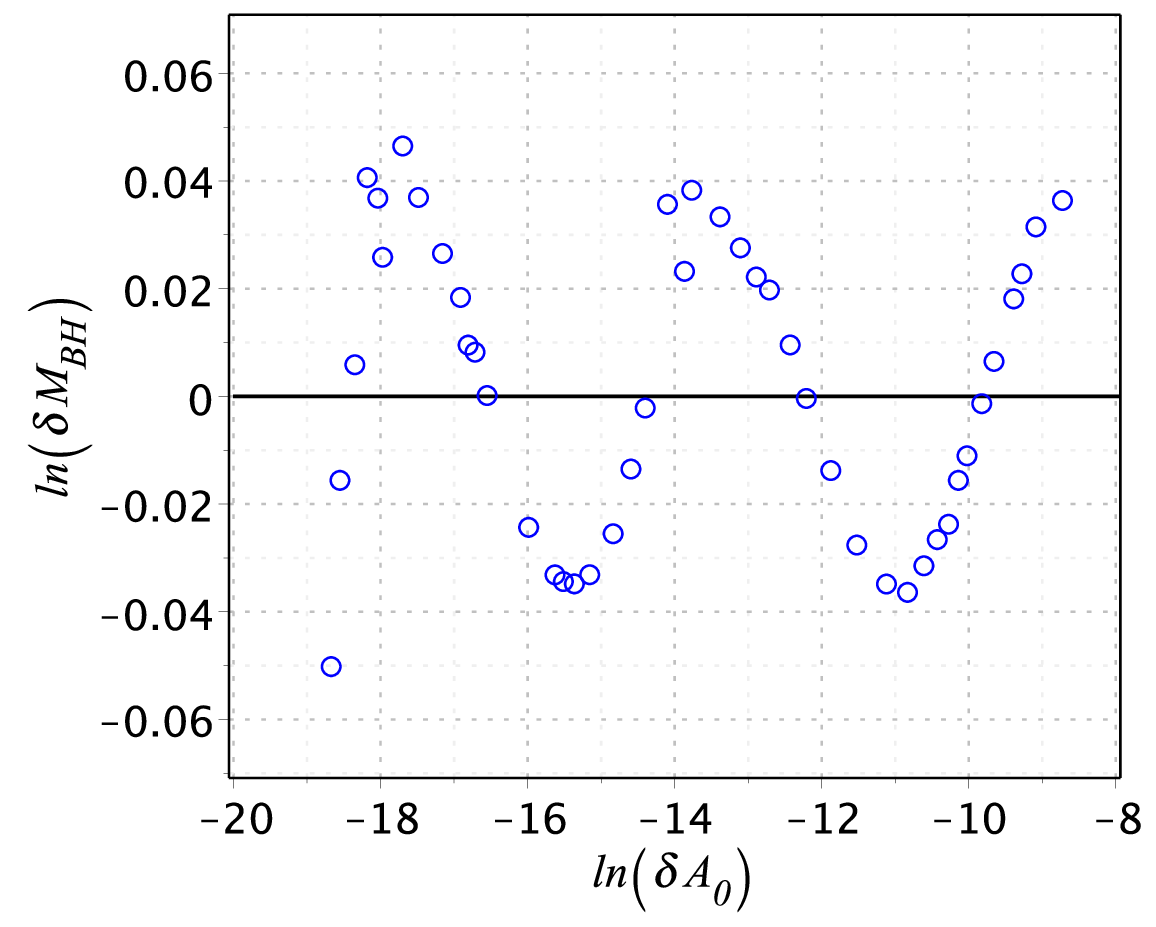}\\
	\begin{center}{\vspace{-1.cm} (a) \vspace{0.5cm}}\end{center}
\includegraphics[width=5.5cm,height=4.5cm]{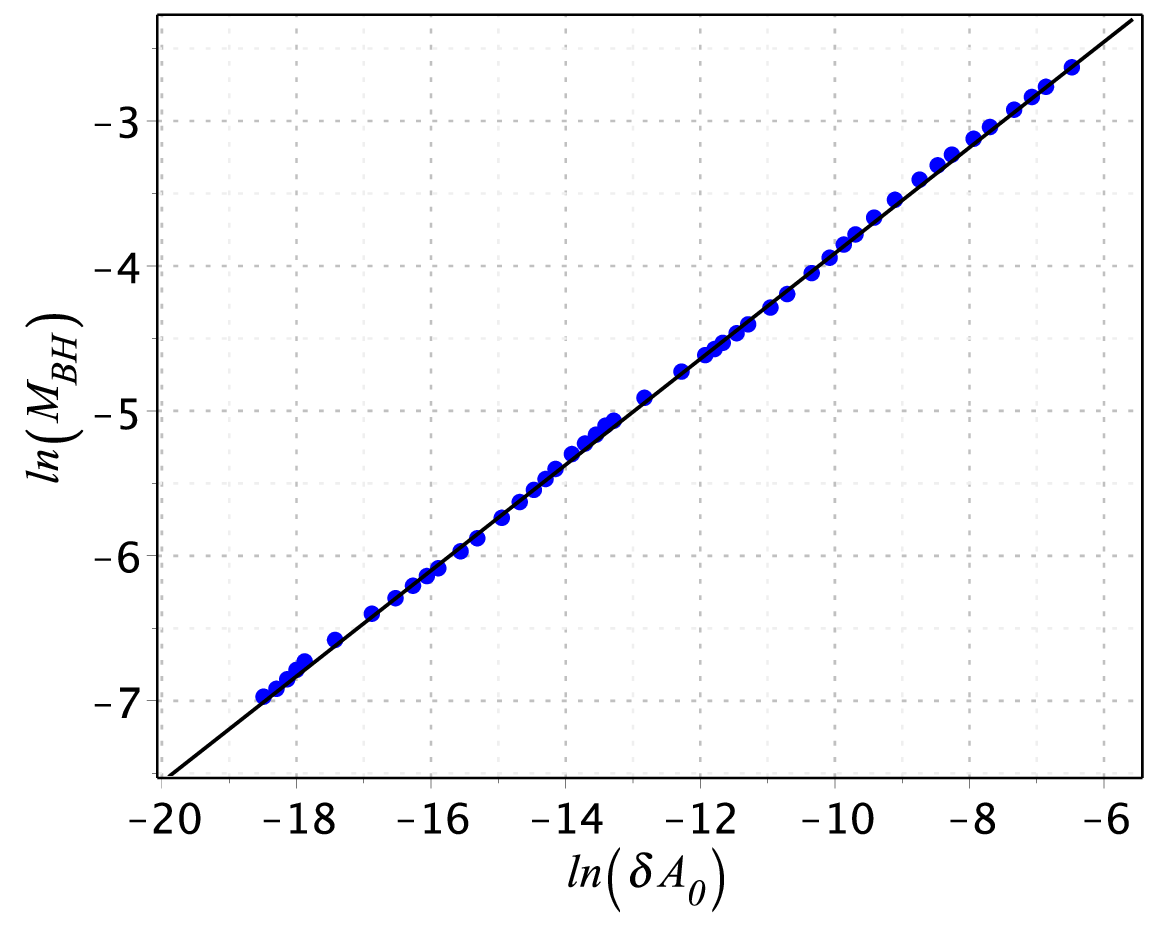}\hspace{0.5cm}
\includegraphics[width=6cm,height=4.45cm]{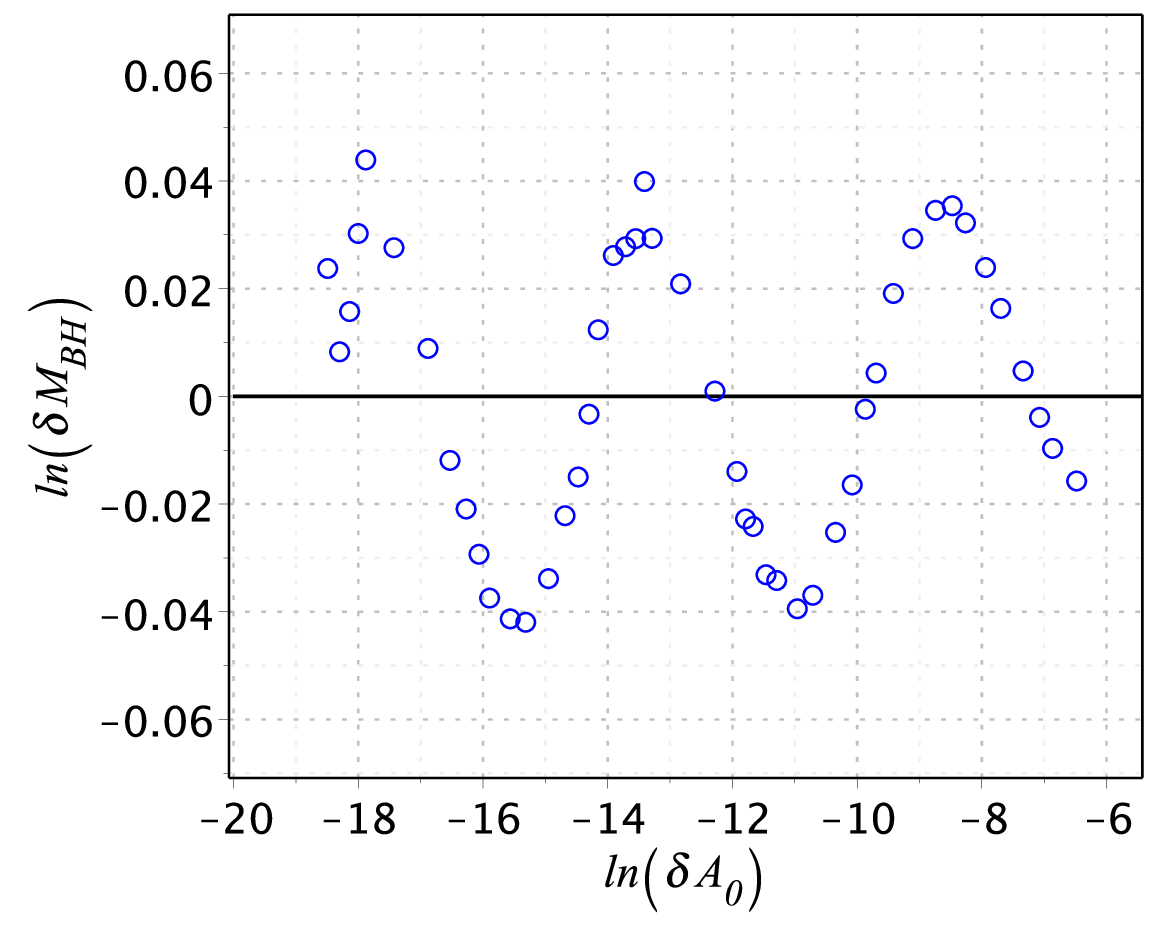}\\
\begin{center}{\vspace{-1.cm} (b) \vspace{0.5cm}}\end{center}
\caption{(a) Two subdomains with $N_1=N_2=500$ as the scalar field truncation orders and interface 
at $r^{(1)}=L_0=0.5 \,\,(x^{(1)}=0)$. (b) Two subdomains with $N_1=N_2=200$, interface at
$r^{(0)}=L_0/3\,\,(x^{(1)}=-0.5)$ and $L_0-0.3$. Here $\delta M_{BH} = M_{BH} - \kappa (\delta A_0)^\gamma$ is the oscillatory component.}
\end{figure*}
\begin{figure*}[h!]
	\includegraphics[width=5.5cm,height=4.5cm]{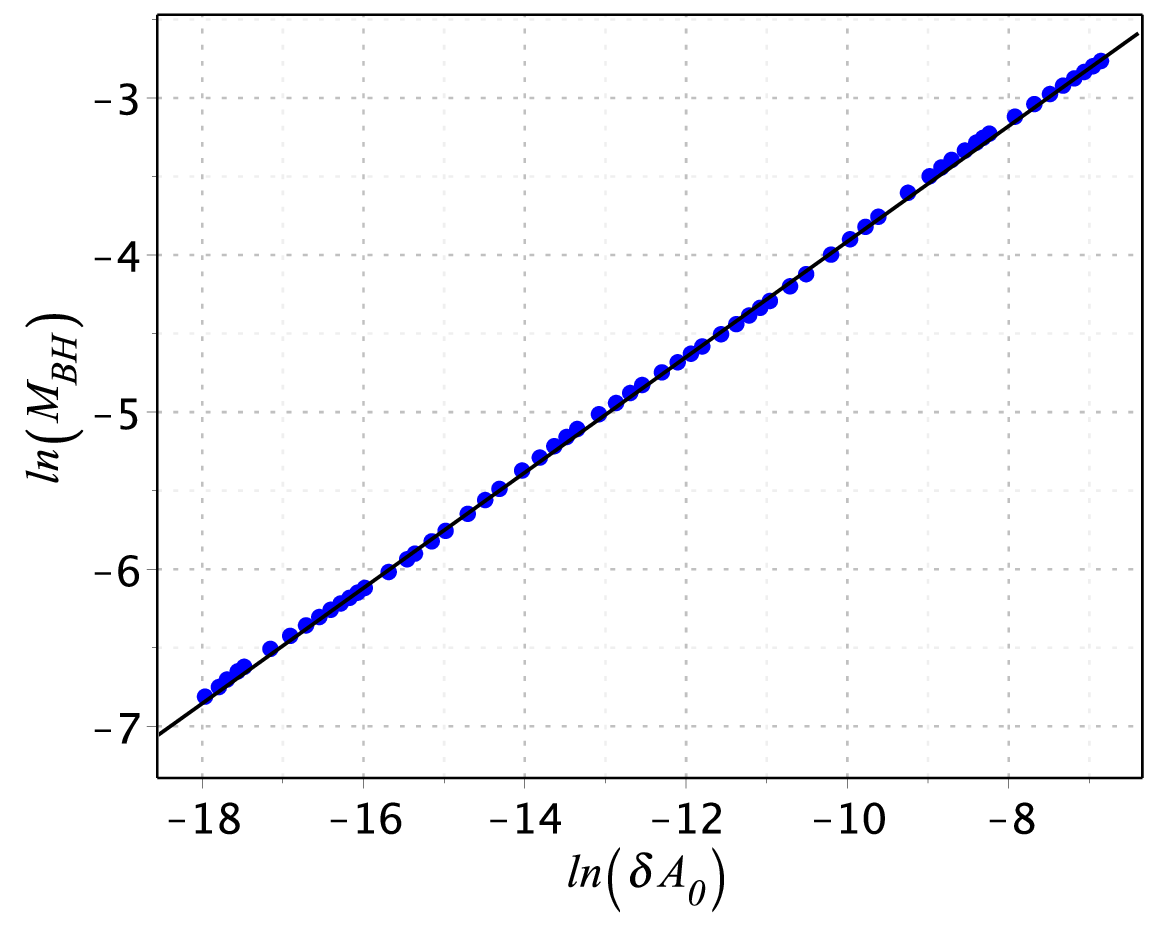}\includegraphics[width=6cm,height=4.45cm]{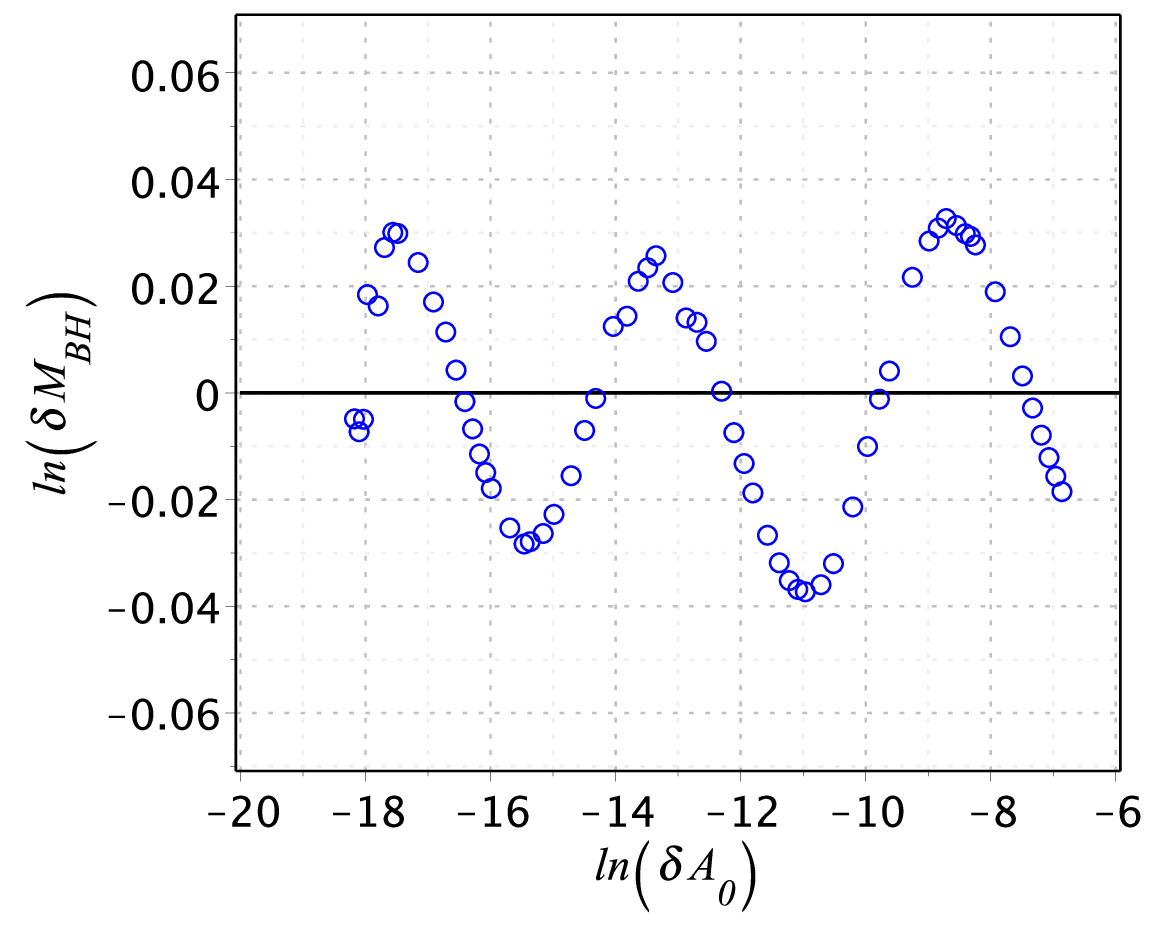}\\
	\begin{center}{\vspace{-1.0cm} (a) \vspace{0.5cm}}\end{center}
	\includegraphics[width=5.5cm,height=4.5cm]{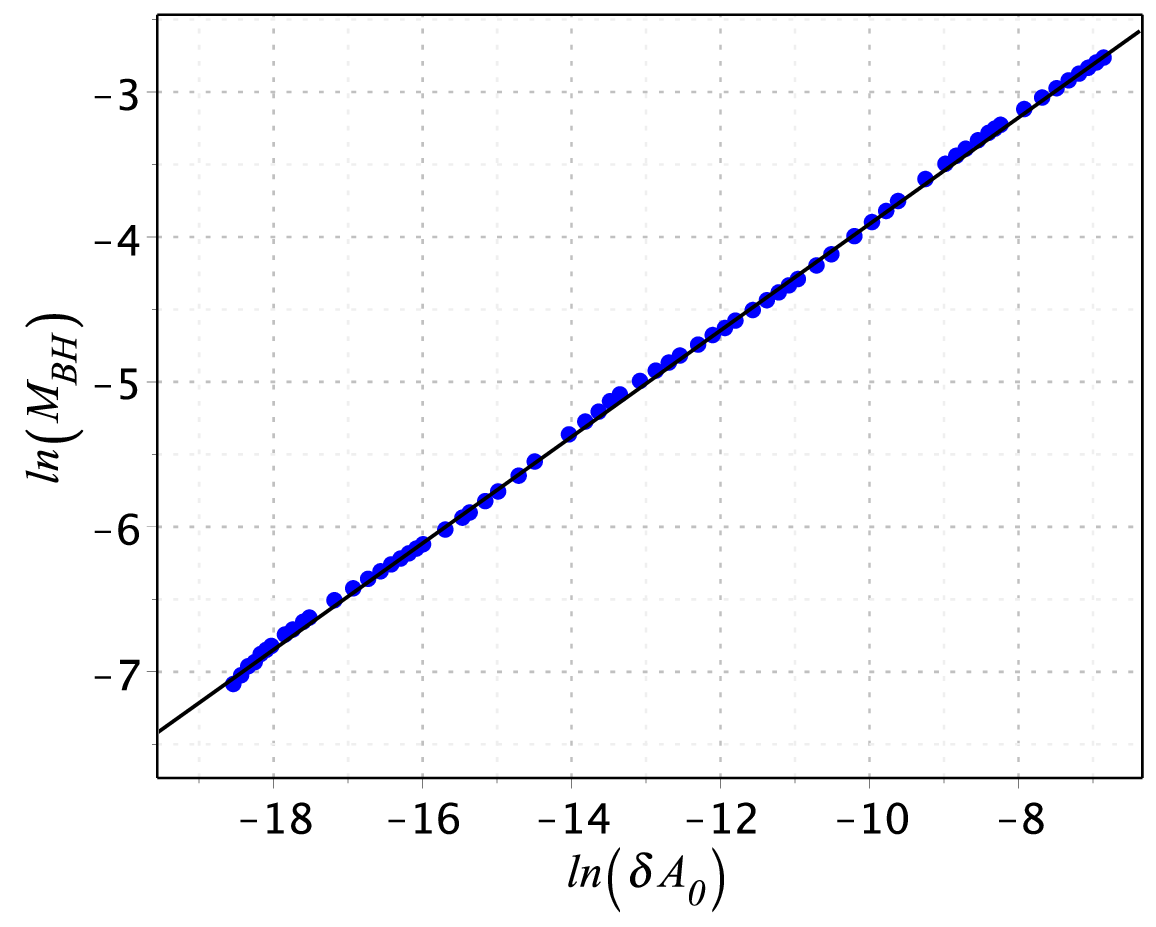}\includegraphics[width=6cm,height=4.45cm]{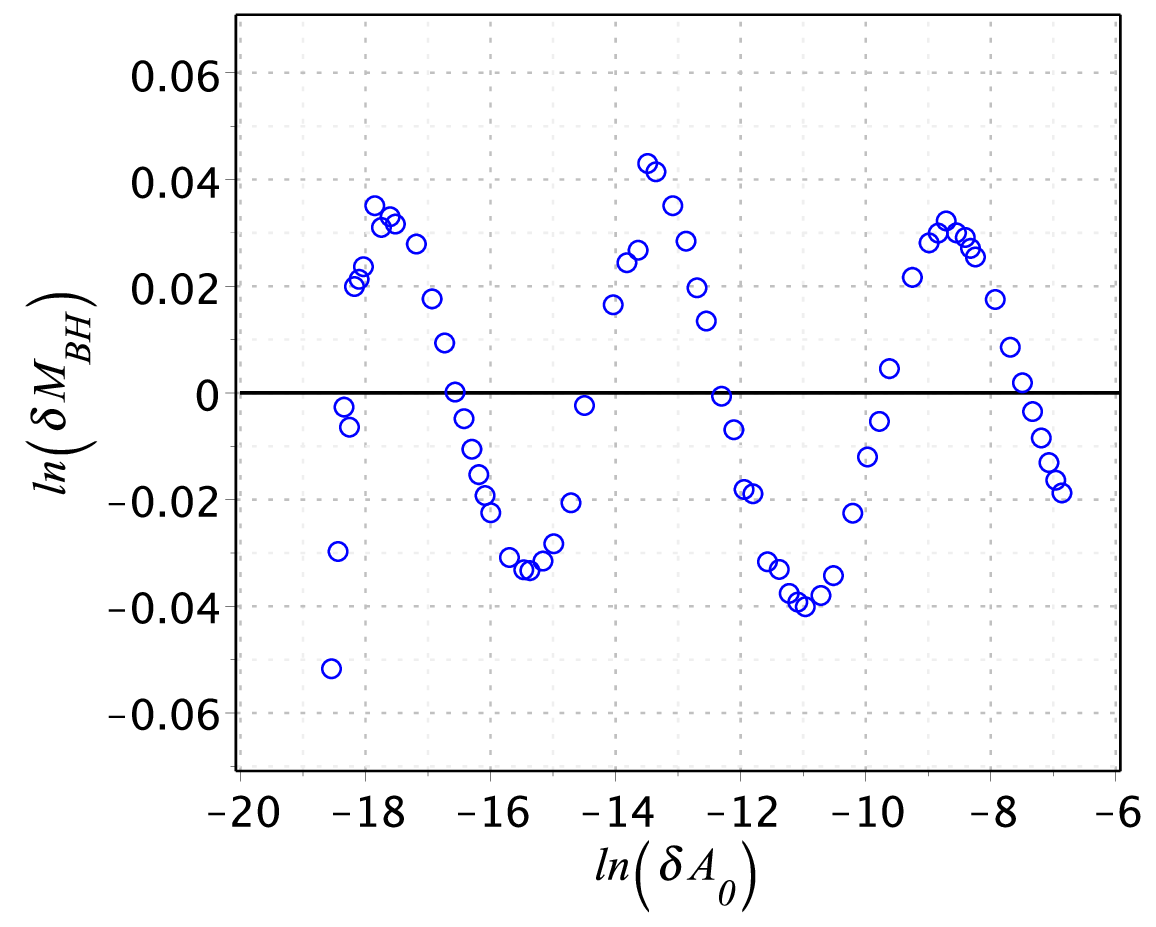}\\
	\begin{center}{\vspace{-1.cm} (b) \vspace{0.5cm}}\end{center}
	\caption{(a) Four subdomains with $N_1=N_2=N_3=200$ and $N_4=300$, interfaces at $r^{(1)}=L_0/7,\,\,(x^{(1)}=-0.75),\; r^{(2)}=L_0/3\,\,(x^{(1)}=-0.5),\;r^{(3)}=L_0\,\,(x^{(1)}=0)$ with $L_0=0.1$. (b) Four subdomains with $N_1=N_2=N_3=105$ and $N_4=140$, interfaces at $r^{(1)}=L_0/19\,\,(x^{(1)}=-0.9),\; r^{(2)}=L_0/3\,\,(x^{(1)}=-0.5),\;r^{(3)}=L_0\,\,(x^{(1)}=0)$ with $L_0=0.1$. Here $\delta M_{BH} = M_{BH} - \kappa (\delta A_0)^\gamma$ is the oscillatory component.}
\end{figure*}
\begin{figure*}[h!]
	\includegraphics[width=5.5cm,height=4.5cm]{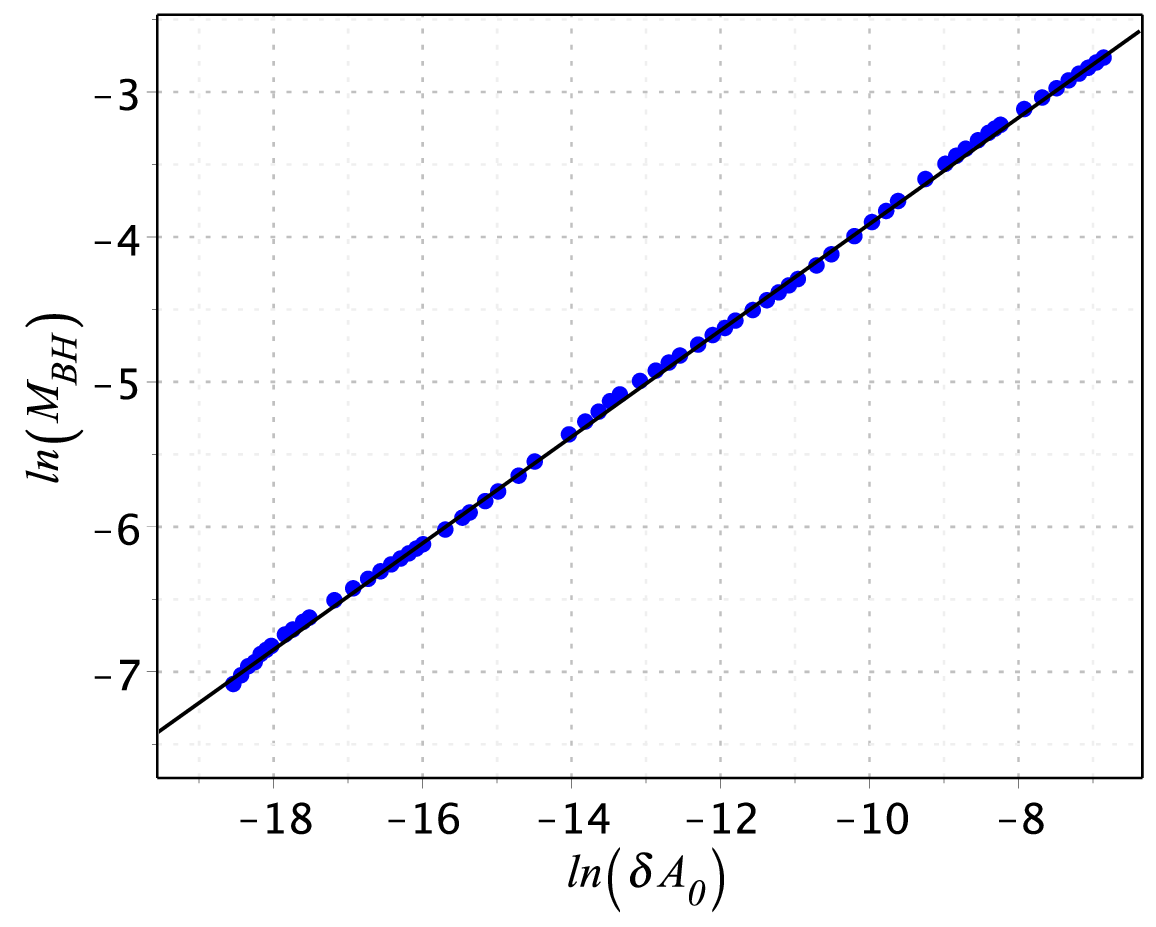}\includegraphics[width=6cm,height=4.45cm]{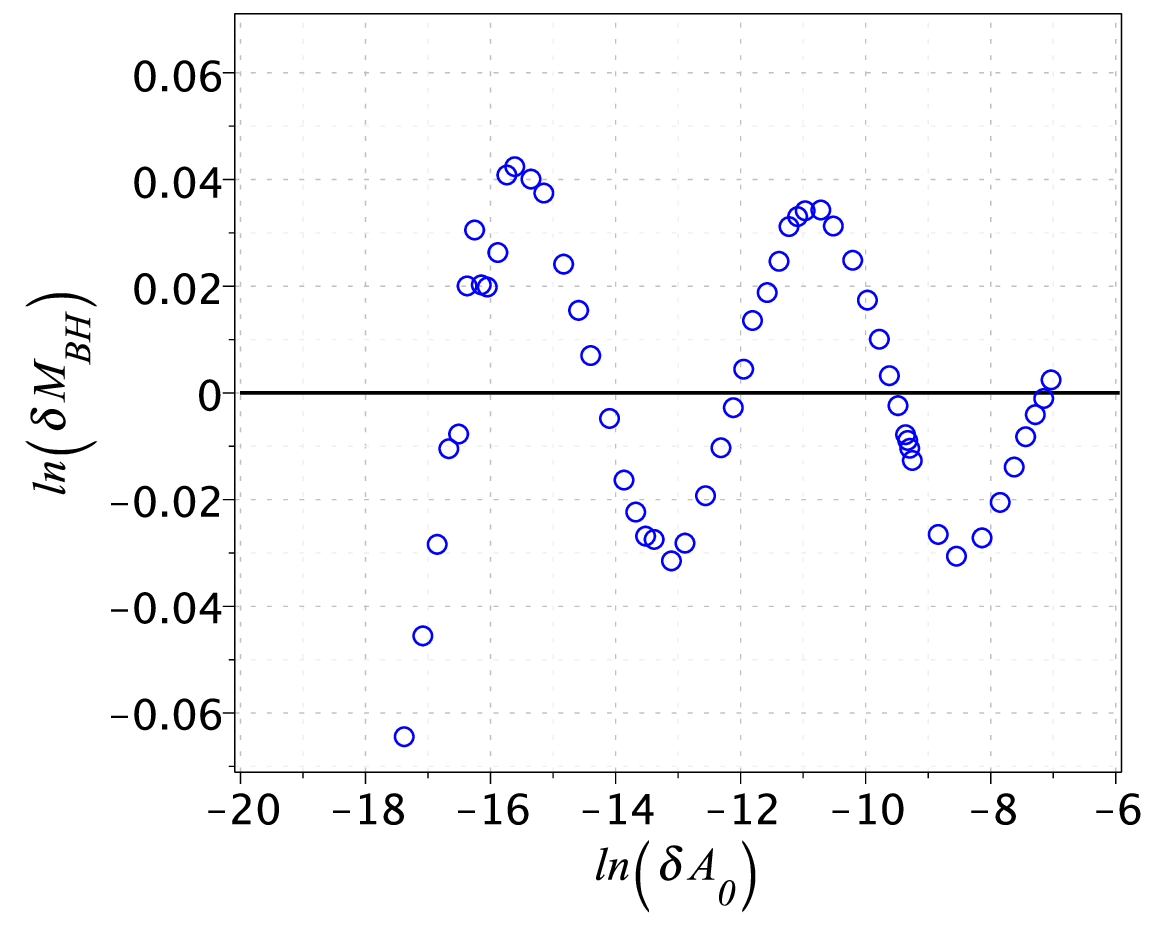}\\
	%\vspace{-0.5cm} %(a)\\
	%\vspace{0.cm}
	\caption{Presence of the potential. Four subdomains with $N_1=N_2=N_3=200$ and $N_4=300$, interfaces at $r^{(1)}=L_0/7,\,\,(x^{(1)}=-0.75),\; r^{(2)}=L_0/3\,\,(x^{(1)}=-0.5),\;r^{(3)}=L_0\,\,(x^{(1)}=0)$ with $L_0=0.1$. Here $\delta M_{BH} = M_{BH} - \kappa (\delta A_0)^\gamma$ is the oscillatory component.}
\end{figure*}

\subsection{Critical collapse}

Critical collapse is still a fertile field of investigation concerning the consequences of the gravitational strong-field regime. Choptuik \cite{choptuik} discovered the essential aspects of what is known by the critical phenomena in gravitational collapse. Following Choptuik's terminology, the space of possible solutions describing the gravitational collapse is divided into supercritical, subcritical, and critical depending on the initial strength of the matter. The supercritical and subcritical solutions enclose those solutions that result in the formation of a black hole and the Minkowski spacetime, respectively. The critical solution located between both has a naked singularity and exhibit discrete self-similarity. For more details, there are thorough reviews of critical collapse \cite{gundlach}.  

We intend to use the multidomain Galerkin-Collocation algorithm to generate the scaling law for the black hole masses corresponding to near-critical solutions. Furthermore, more relevant is to show the scaling law's oscillatory component due to the discrete self-similarity of the critical solution \cite{choptuik,gundlach}.

We start with the initial data (\ref{eq51}) with $A_0$ dictating the initial scalar field amplitude, and as mentioned, we have evolved the system with the adaptive stepsize Cash-Karp integrator \cite{cash-karp}.  Depending on the value of $A_0$, we can obtain subcritical or critical solutions, and between them, there exists a value $A_0^*$ corresponding to the critical solution. In all numerical experiments, we considered near-critical ($A_0 \simeq A_0^*$) and collected the final Bondi mass when the apparent horizon forms. The scaling relation satisfies the following expression \cite{gundlach_97}
\begin{equation}
\ln(M_{AH}) =  \gamma \ln(\delta A_0) + f(\delta A_0) + \kappa \label{eq53}
\end{equation}

\noindent where $\kappa$ is a constant that depends on the initial data family, $\gamma$ is the critical exponent, $f$ is an oscillatory function with period $ \varpi= \Delta/2\gamma$, and $\Delta$ is the echoing period of the DSS critical solution.

To construct the scaling law, we determine the Bondi mass when the apparent horizon forms. Fig. 5 illustrates the Bondi mass decay leaving at the end a minimum value just before the integration diverges ($\beta \rightarrow \infty$). This instant is in agreement with the formation of the step function of $\mathrm{e}^{2 (\beta-H)}$ as established by Christodoulou \cite{christoudolou} prior to the formation of the apparent horizon.

In the sequence, we present the black hole scaling law and the oscillatory component obtained after subtracting the term $ \gamma \ln(\delta A_0)$ of Eq. (\ref{eq53}). For this task, we obtained numerical data using algorithms with a distinct number of subdomains. We concentrate more collocation points near the origin by choosing the first interface location close to $x=-1\,\, (r=0)$ in the intermediate computational subdomain in all of the code versions. 

We present in Fig. 6 the results using two versions of the two subdomains. In the first version, Fig. 6(a), we have set $N_1=N_2=500$, $x^{(1)}=0$ meaning that the interface in the physical domain is $r^{(1)}=L_0=0.5$. For the plots in Fig. 6(b), we have set $N_1=200$, $N_2=200$ but $x^{(1)}=-0.5$ implying in $r^{(1)}=L_0/3$ and we have set $L_0=0.3$. It becomes clear the effect of concentrating more collocation points near the origin, since using a smaller total truncation order for the scalar field (1000 versus 400 collocation points) we obtained similar results. We recall that smaller Bondi masses correspond to the apparent horizons forming close to the origin.  

The critical exponent and the period of the oscillatory component are determined from the numerical data of Fig. 6. We obtained, after the best fit of Eq. (\ref{eq53}) the following parameters for the Figs. 6(a) and 6(b), respectively 
\begin{eqnarray}
\gamma \approx 0.3659,\;\; \varpi \approx 4.292. \\
\nonumber \\
\gamma \approx 0.3647,\;\; \varpi \approx 4.779.
\end{eqnarray}

\noindent The perturbation analysis relates the echoing period of the critical solution, $\Delta$ with the above parameters through $\Delta=2 \gamma \varpi$. With the present values, we have found $\Delta \approx 3.139$ and $\Delta \approx 3.468$ representing a discrepancy of $4.61\%$ and $1.34\%$, respectively if compared with $\Delta \approx 3.44$ obtained from the perturbation approach to the full accuracy \cite{gundlach_97}. 

In the next set of numerical experiments, we have considered the four subdomains algorithms with distinct interface locations and truncation orders. The plots presented in Fig. 7(a) correspond to the interfaces locations at $r^{(1)}=L_0/7, r^{(2)}=L_0/3, r^{(3)}=L_0$  and scalar field truncation orders $N_1=N_2=N_3=200,N_3=300$. In the plots presented in Fig. 7(b) the interface locations are $r^{(1)}=L_0/19, r^{(2)}=L_0/3, r^{(3)}=L_0$ with the truncation orders $N_1=N_2=N_3=105,N_3=140$.  In both cases, we have selected $L_0=0.1$. 

Despite the distinct collocation points in each case, the outcomes are similar. From the numerical data of Figs. 7(a) and 7(b), we can extract the following values for the critical exponent and period of the oscillatory component:
\begin{eqnarray}
\gamma \approx 0.36767,\;\; \varpi \approx 4.540. \\
\nonumber \\
\gamma \approx 0.3670,\;\; \varpi \approx 4.549,
\end{eqnarray}

\noindent implying in $\Delta \approx 3.338$ and $3.339$ and discrepancies of about $2.96\%$ and $2.93\%$, respectively if compared with the linear perturbation value $\Delta \approx 3.44$.

As the final application, we have considered introducing the potential $U(\phi) = \lambda \phi^4/4$ and the four subdomain algorithm with $N_1=N_2=N_3=200, N_4=300$. We chose, without loss of generality, $\lambda=1$ and proceeded the Bondi mass's determination when the apparent horizon forms.  The results are shown in Fig. 8 with the scaling law and the oscillatory component and the corresponding parameters are
\begin{eqnarray}
\gamma \approx 0.36852,\;\; \varpi \approx 4.636. 
\end{eqnarray}

\noindent We obtained $\Delta \approx 3.417$ very close to the actual predicted value of $3.44$.

\section{Summary and future perspectives}

The present work is the first of the systematic development of the multidomain or domain decomposition technique connected with the Galerkin-Collocation method and applied to situations of interest in numerical relativity.  We have considered the self-gravitating and spherically symmetric scalar field in the characteristic scheme as the first work. Nevertheless, some other works applied spectral codes to study critical collapse \cite{edu_oliv,kidder_19}. In particular, the last reference deals with the critical collapse in 3D.

We remark how we have compactified and divided the spatial domain $\mathcal{D}: 0 \leq r < 0$ into several subdomains. We have introduced first an intermediate computational domain $-1 \leq x \leq 1$ using an algebraic map given by (\ref{eq18}), and with linear mappings, we define several subdomains labeled by $-1 \leq \xi^{(l)} \leq 1,\;\;l=1,2,..,n$ as shown by Fig. 1. This scheme is the backbone for implementing the algorithm.

In the sequence, we establish the basis functions necessary for the spectral approximations of the metric functions and the scalar field in each subdomain. We have adopted the patching method for joining the solutions in each subdomain through the transmission conditions in the subdomains' interfaces. We have called attention that the transmission conditions for the Klein-Gordon equation are not unique, but we opted for the simplest form given by Eqs. (\ref{eq24}) and (\ref{eq25}). No penalty techniques are considered here, rendering our approach more simple and easy to implement. 

Another additional feature is the use of two distinct sets of collocation points in each subdomain inspired in the characteristic scheme's hierarchy. The first set is for the metric functions $\beta(u,r)$ and $V(u,r)$ and the second set for the scalar field $\Phi(u,r)$, where in general there are more points in the first set.  

We provided the validation of the Galerkin-Collocation algorithm from two numerical tests. The first was the verification of the Newman-Penrose constant \cite{NP}, and the second test consisted of verifying the Bondi formula (\ref{eq9}). We considered an initial data that, despite not forming an apparent horizon has an initial amplitude close to it. In general, the results exhibited exponential convergence. However, the convergence depends on several factors: the number of subdomains, the place of the interfaces, the number of collocation points in each subdomain, and the map parameter $L_0$. 

We have reproduced the Choptuik's scaling law and the corresponding oscillatory component related to the critical collapse as a valid application.  We have tested the code with two and four subdomains together with less than 1,000 total collocation points distributed in all subdomains. Furthermore, we have used the value of the Bondi mass evaluated when the apparent horizon forms. In all cases, the critical exponent is slightly different ($\approx 1.6\%$ of relative deviation) from
 those obtained by Choptuik \cite{choptuik}, Hod and Piran \cite{piran}, and Purrer at al. \cite{purrer}. Also, the values of the echoing period present in a slight discrepancy (around $1 \%$). However, the motivation was to present reliable results with a moderate resolution.  We believe the present results show the domain decomposition GC method's effectiveness for the well-known and computationally expensive gravitational critical collapse and its fine structure.

The domain decomposition GC method can be extended to other contexts and higher spatial dimensions. In this direction, we are currently implementing the Galerkin-Collocation method in some cases of interest.  We are currently preparing the code for the problem of the self-gravitating spherically symmetric scalar field in the Cauchy formulation. Also, we are implementing the GC domain decomposition for the axisymmetric Bondi problem. Finally, the dynamics using the BSSN formulation is one of our primary objectives since we would tackle spherical and non-spherical systems. 
%\newpage
\begin{acknowledgements}
	M. A. acknowledges the financial support of the Brazilian agency Coordena\c c\~ao de Aperfei\c coamento de Pessoal de N\'\i vel Superior (CAPES). W. O. B. thanks to the 
Departamento de Apoio \`a Produ\c c\~ao Cient\'\i fica e Tecnol\'ogica (DEPESQ) for the financial support, also to the Departamento de F\'\i sica Te\'orica for the hospitality, both at the Universidade do Estado do Rio de Janeiro. H. P. O. thanks Conselho Nacional de Desenvolvimento Cient\'\i fico e Tecnol\'ogico (CNPq) and Funda\c c\~ao Carlos Chagas Filho de Amparo \`a Pesquisa do Estado do Rio de Janeiro (FAPERJ) (Grant No. E-26/202.998/518 2016 Bolsas de Bancada de Projetos (BBP)).
\end{acknowledgements}

%\appendix*

\section*{Appendix: Basis functions}

We present the basis functions used in the spectral approximations of $\Phi, \beta $, and $ V $ established in the first and last subdomains.  The basis functions for the scalar field $\Phi$ (cf. Eq. (\ref{eq17}))is
\begin{eqnarray}
\psi_k^{(1)}(r) &=& \frac{1}{2}\left(TL_{k+1}^{(1)}(r) + TL_{k}^{(1)}(r)\right),
\end{eqnarray}

\noindent where $TL^{(1)}_k(r)$ with $k=0,1,..,N_1$ are the rational Chebyshev polynomials defined in the first subdomain. The basis for  the metric function $\beta$ is
\begin{eqnarray}
\chi_k^{(1)}(r) &=& \frac{1}{8}\bigg[\frac{(1+2k)}{3+2k}TL_{k+2}^{(1)}(r) +\frac{4(1+k)}{3+2k}TL_{k+1}^{(1)}(r) \nonumber \\
& & + TL_{k}^{(1)}(r)\bigg].
\end{eqnarray}

\noindent Each of the above basis functions satisfies the condition $(\ref{eq11})$. For the last subdomain, we have chosen the corresponding rational Chebyshev function as the basis for all spectral approximations of $\Phi, \beta $ and $ V $.


\begin{thebibliography}{99}

\bibitem{grand_novak} Grandcl\'ement, P., Novak, J.: Living Rev. Relativ. {\bf 12}, 1 (2009)

\bibitem{canuto_88} Canuto, C., Quarteroni, A., Hussaini, M. Y., Zang, T. A.: \textit{Spectral Methods in Fluid Dynamics}, 
	Springer-Verlag (1988)

\bibitem{gottlieb_01} Gottlieb, D. Hesthaven, J. S.: J. Comp. App. Math. - Special issue on Numerical Analysis, Vol. VII: Partial Differential Equations, 128, 1 - 2, 83 (2001)

\bibitem{canuto_new} Canuto, C., Quarteroni, A., Hussaini, M. Y., Zang, T. A.: \textit{Spectral Methods - Evolution 
	to Complex Geometries and Applications to Fluid Dynamic}, Springer-Verlag (2007)

\bibitem{orszag_80} Orszag, S.: J. Comp. Phys. {\bf 37}, 70 (1980)

\bibitem{kopriva_86} Kopriva, D. A.: Appl. Numer. Math. {\bf 2}, 221 (1986)

\bibitem{kopriva_89} Kopriva, D. A.: SIAM J. Sci. Sta. Comp. {\bf 10}, 1 120 (1989)

\bibitem{faccioli_96} Faccioli, E., Maggio, F., Quarteroni, A., Tagliani, A.: Geophysics {\bf 61}, 1160 (1996)

\bibitem{bona} Bonazzola, S., Gourgoulhon, E., Salgado, M., Marck, J. A.: Astron. Astrophys. {\bf 278}, 421 (1993)

\bibitem{pfeifer} Pfeifer, H. P.: \textit{Initial data for black hole evolutions}, Ph.D. thesis,  arXiv:gr-qc/0510016 (2003)

\bibitem{ansorg} Ansorg, M.: Class. Quantum Grav. {\bf 24}, S1 (2007)

\bibitem{spec} Spectral Einstein Code, https://www.black-holes.org/code/SpEC.html

\bibitem{lorene} LORENE (Langage Objet pour la Relativit\'e Num\'erique), http://www.lorene.obspm.fr

\bibitem{kidder_00} Kidder, L. E., Scheel, M. A., Teukolsky, S. A.: Phys. Rev. D {\bf 62}, 084032 (2000)

\bibitem{szilagyi_09} Szil\'agyi, B., Lindblom, L., Scheel, M. A.: Phys. Rev. D {\bf 80}, 124010 (2009)

\bibitem{Hemberger_13} Hemberger, D. A., Scheel, M. A., Kidder, L. E., Szil\'agyi, B., Lovelace, G., Taylor, N. W., Teukolsky, S. A.:
	Class.  Quantum Grav. {\bf 30}, 115001 (2013)

\bibitem{sxs_col} Boyle, M. et al.: \textit{The SXS Collaboration catalog of binary black hole simulations}, arXiv: 1904.04831 (2019)

\bibitem{winicour_12} Winicour, J.: Living Rev. Relativity {\bf 15}, 2 (2012)

\bibitem{oliveira_14} de Oliveira, H. P.,  Rodrigues, E. L.: Phys. Rev. D {\bf 90}, 124027 (2014)

\bibitem{barreto_18} Barreto, W., Clemente, P. C. M., de Oliveira, H. P.: Phys. Rev. D {\bf 97}, 104035 (2018)

\bibitem{barreto_18_2} Barreto, W., Clemente, P. C. M., de Oliveira, H. P., Rodrigue-Mueller, B.: Gen. Rel. Grav. {\bf} 50, 71 (2018)

\bibitem{barreto_19} Barreto, W., Crespo, J. A., de Oliveira, H. P., Rodrigues, E. L.: Class. Quant. Grav. {\bf 36}, 215011 (2019)

\bibitem{crespo_19} Crespo, J. A., de Oliveira, H. P., Winicour, J.: Phys. Rev. D {\bf 100}, 104017 (2019)

\bibitem{ADM} Arnowitt, R., Deser, S., Misner, C. W.: \textit{Republication of: The dynamics of general relativity}. Gen.
	Relativ. Gravit. {\bf 40}, 1997 (2008)

\bibitem{bondi} van der Burg, M., Bondi, H.,  Metzner, A.: Proc. R. Soc. London Ser. A {\bf 269}, 21 (1962)

\bibitem{gomez_jmp} G\'omez, R., Winicour, J.: J. Math. Phys. {\bf 33}, 1445 (1992)

\bibitem{boyd} J. P. Boyd, {Chebyshev and Fourier Spectral Methods} (Dover Publications, New York, 2001).

\bibitem{peyret} Peyret, R.: \textit{Spectral methods for incompressible viscous flow}, Springer-Verlag, New York (2002)

\bibitem{choptuik} Choptuik, M. W.: Phys. Rev. Letters {\bf 70}, 9 (1993)
	
\bibitem{cash-karp} Cash, J. R.,  Karp, H. A.: \textit{Variable order Runge-Kutta method for initial value problems 
	with rapidly varying righthand sides}, ACM Transactions on Mathematical Software {\bf 16}, 201 (1990)

\bibitem{NP} Newman, E. T., Penrose, R.: Proc. R. Soc. London Ser. A {\bf 305}, 175 (1968)

\bibitem{NC} Abramowitz, M., Stegun, I.A.: \textit{Handbook of Mathematical Functions with Formulas, Graphs, and Mathematical Tables}. 
	New York: Dover (1972)

\bibitem{gundlach} Gundlach, C. Martin-Garcia, J. M.: Living Rev. Relativ. {\bf 10}, 5 (2007)

\bibitem{gundlach_97} Gundlach, C.: Phys. Rev. D {\bf 55}, 695 (1997)

\bibitem{christoudolou} Christodoulou, D.: Commun. Math. Phys. {\bf 109}, 613 (1987)

\bibitem{edu_oliv} Rodrigues, E. L., de Oliveira, H. P.: Phys. Rev. D {\bf 82}, 104023 (2010)

\bibitem{kidder_19} Deppe, N., Kidder, L. E.,Scheel, M. A., Teukolsky, S. A.:Phys. Rev. D {\bf 99}, 024018 (2019)

\bibitem{piran} Hod, S. Piran, T.: Phys. Rev. D {\bf 55}, 440 (1997)

\bibitem{purrer} Purrer, M., Husa, S., Aichelburg, P. C.: Phys. Rev. D {\bf 71}, 104005 (2005)

%\bibitem{barreto_17} Barreto, W., de Oliveira, H. P., Rodriguez-Mueller, B.: Gen. Rel. Grav. {\bf 49}, 107 (2017)


\end{thebibliography}
\end{document}